\documentclass{aa}
\usepackage{psfig}
\usepackage{lscape}
\usepackage{amsmath}
\usepackage[figuresleft]{rotating}

\newcommand{\MC}{\multicolumn}
\newcommand{\kms}{km~s$^\mathrm{-1}$}
\newcounter{qub}
\newcommand{\qq}{\addtocounter{qub}{1}\arabic{qub}}

\begin{document}

\title{Environment status of blue compact galaxies and
  trigger of star  formation
}

\author{Simon A. Pustilnik\inst{1,2} \and
Alexei Y. Kniazev\inst{1,2} \and
Valentin A. Lipovetsky\inst{1}\thanks{Deceased 1996 September 22.}  \and
Andrei A. Ugryumov\inst{1,2}}

\offprints{S. Pustilnik, \email{sap@sao.ru}}

\institute{
Special Astrophysical Observatory RAS, Nizhnij Arkhyz,
Karachai-Circassia,  369167 Russia  \\
\and Isaac Newton Institute of Chile, SAO Branch
}

 \date{Received  26 December, 2000; accepted  26 March, 2001}

 \abstract{
The work studies\thanks{Tables 2, 3 and A1 with their notes and Figures A1--A3
are only  available in the electronic form via http://www.edpsciences.org}
of the environment of low-mass galaxies with
active star formation (SF) and a possible trigger of SF bursts due to
gravitational interaction. Following the study by Taylor et al. (1995), we
extend the search for possible disturbing galaxies of various masses
to a much larger sample of 86 BCGs from the sky region of the Second Byurakan
survey (SBS). The BCG magnitudes and radial velocities are revised and
up-dated. The sample under study is separated by the criteria:
EW([O{\sc iii}]$_{\lambda5007}$) $>$ 45~\AA\ and V$_\mathrm{h} <$ 6,000~\kms\
and should be representative of all low-mass galaxies which
experience SF bursts. We argue that the moderate tidal disturbers should be
taken into account, and incorporate the respective range of distances in the
search for disturbing neighbours. The majority of the neighbours in the
vicinity of the studied BCGs are found through the study of their
environment among UZC (Falco et al.~\cite{Falco99}) galaxies, and the
follow-up careful search of the fainter galaxies in the
NED database. For the remaining BCGs, the neighbouring galaxies are found
based on the results of the SAO 6\,m telescope spectroscopy.
By studing
the data on the radial velocities of galaxies in the vicinity of BCGs we
found: 1) 33 of the studied BCGs ($\sim$38.5\%) are associated with
significantly brighter galaxies ($\Delta B \geq$ 1.5$^\mathrm{m}$);
2) 23 BCGs ($\sim$26.5\%) have neighbours either of comparable or significantly
lower brightness; 3) 14 of the studied BCGs (16\%) with no evident
associated galaxy are either certain, or probable, mergers.
Summarizing, we conclude that in $\sim$80\% (or more) BCGs from the studied
sample, the SF bursts are triggered either by tidal action of various
strengths
from other galaxies, or due to mergers of low-mass galaxies. We briefly
discuss the implications of our main conclusion for evolutionary
links of BCGs to other types of low-mass galaxies.
Part of our sample falls into the volume belonging to the Local
Supercluster.
Therefore we formulate the results separately on the `Local Supercluster
volume' and  `general field region'.
The total fractions of BCGs likely triggered by interaction with other
galaxy are respectively, $\sim$84.5\% and 80\% for the nearby volume and for
the general field.
The fractions of BCGs with significantly brighter disturbers in these
two groups are seemingly different ($\sim$54$\pm$14\% vs $\sim$31.5$\pm$7\%,
respectively).
Among the so called ``isolated'' BCGs (that is, without  a bright
neighbouring galaxy) in both the Local Supercluster volume and in general
field, $\sim$43$\pm$10\% are probably disturbed by dwarf galaxies and
$\sim$26$\pm$8\%  have  a merger morphology.
In the Appendix we present the results of the spectroscopy with the
SAO 6\,m
telescope of 27 galaxies
in  an attempt to find possible disturbing galaxies in the
vicinity of some of the sample BCGs.
   }

\authorrunning{S.A. Pustilnik et al.}

\titlerunning{Environment Status of BCGs}

   \maketitle

\section{Introduction}

Blue compact galaxies (BCGs) with strong emission lines or
\ion{H}{ii}-galaxies are low-mass objects with the current star-formation
(SF) rate exceeding many times this parameter
averaged over cosmological time.
The observed SF rates for most BCGs are so high that star formation could be
sustained at the current level only on a timescale significantly lower
than 1/H$_\mathrm{0}$\footnote{H$_\mathrm{0}$ is the Hubble constant,
accepted hereafter as 75 \kms~Mpc$^\mathrm{-1}$.}
before their neutral gas reservoir will be completely depleted
(e.g., Thuan~\cite{Thuan91}).
This implies that either these objects are rather young, with typical ages of
less than one or a few Gyr, or SF in BCGs is highly variable and proceeds in
short episodes (bursts) with a typical duration of 10--100 Myr, intermittent
with long periods of SF activity at levels much lower than that during
the SF burst. In the course of such SF bursts,  BCGs experience
significant
brightening relative to their quiescent state -- an average of 0\fm75 in the
small BCG sample from Papaderos et al.~(\cite{Papa96}), and up-to
$\sim$2\fm0 in a large sample from Lipovetsky et al.~(\cite{Lipovet}).
Therefore being in general low-mass and underluminous systems with a
wide range of baryon (stars+gas) mass
(10$^\mathrm{7} < M_\mathrm{bar} < $10$^\mathrm{10}$ M$_{\odot}$),
BCGs sometimes can be as bright as M$_\mathrm{B}=-$20\fm0.

In the great majority of properly
studied BCGs, the traces of old low-mass populations are detected
(e.g., Thuan~\cite{Thuan83}). This implies that, in general, BCGs are rather
old objects, and their evolution on
the Hubble time scale can effectively be determined by short, but very
intensive SF, bursts. The option of truly young galaxies is
still probable for very small fraction of BCGs ($\sim$1\%) with extremely
low metallicities (e.g., Izotov \& Thuan~\cite{IT99}; Kniazev et
al.~\cite{Kniazev00}; Pustilnik et al.~\cite{Pustilnik01a}).

While the mechanisms that regulate and terminate SF bursts on the ``short''
time scales are more or less known and partly understood (SNe and stellar
winds), the nature of the mechanisms which trigger the collapse of gas clouds
and the onset of SF bursts in low-mass galaxies has been debated since the
early 70-s.

A general approach to the question should take into account that even though
BCGs do not comprise a homogeneous class of objects (since they
possess a wide range of observed morphologies), to a first approximation they
all are gas-rich galaxies with high enough specific angular momentum, and
therefore more or less flattened rotating "disks" (e.g. Taylor et al.
\cite{Taylor93,Taylor95}; Chengalur et al.~\cite{Chengalur95};
van Zee et al. \cite{Zee98}).  In the least massive BCGs,
the rotation energy is not much greater than that of random motion, and their
"disks" are thicker. On the other hand the  BCGs similar to Mkn~996 that is,
with the light distribution typical of elliptical galaxies
are quite rare (Doublier et al.~\cite{Doublier97}).
The latter is probably indicative of their post-merger nature (Thuan et
al.~\cite{Thuan96}).

The concept of equilibrium and gravitational instability in a rotating disk
is based on the local stability criterion, introduced by
Toomre~(\cite{Toomre64}).
There is a threshold surface density (depending on the local epicyclic
frequency and the dispersion of the random velocity component
(Toomre~\cite{Toomre64}; Kennicutt~\cite{Kennicutt89})), above which gas is
unstable.
Positions of SF regions in disk galaxies, examined on a variety of galaxy
types (Sa--Sm, Im, LSB) are consistent with the criterion of Toomre--Kennicutt
(Kennicutt~\cite{Kennicutt89}; van der Hulst et al.~\cite{Hulst93}).
Positions of SF regions in BCG/\ion{H}{ii} galaxies also fulfill this
criterion (e.g., Taylor et al.~\cite{Taylor94}; van Zee et al.~\cite{Zee98}).

On the other hand, most of the observed cases of enhanced SF in {\it massive}
gas-rich galaxies are associated with various interactions from other
galaxies.
Keel~({\cite{Keel93}), from the kinematical study of a well-selected sample of
interacting galaxies has drawn the important conclusion that gravitational
instability in their disks is driven initially by external perturbations
(see also Bernloehr~\cite{Bernloehr93}).

At a first glance this mechanism could play an important role for BCGs as
well. However, the studies of the spatial distribution of BCGs indicate that
only a minority are tightly connected to massive galaxies or their
systems (e.g.,  Salzer~\cite{Salzer89}; Campos-Aguilar \&
Moles~\cite{Campos91}; Vilchez~\cite{Vilchez93}; Pustilnik et
al.~\cite{Pustilnik95}).

This fact does not imply that interactions are of low significance
as SF triggers in BCGs. However  many researchers seem
to  favor some internal process as the main trigger of enhanced SF.
In particular, the mechanism  suggested by Gerola et al.
(\cite{Gerola80}),  which is based on the stochastic collisions and merging
of large gas clouds is noted. Another type of internal trigger
is connected with the hypothesis of cyclic gas re-processing, with the
characteristic time between the subsequent SF bursts related to the fall-off
back to the "disk" of the gas lifted to the halo during the previous SF burst
(Salzer \& Norton~\cite{Salzer99}).

The two alternative models have been suggested to overcome the apparent
difficulties of the hypothesis in explaining the interaction-induced SF burst
in BCGs, as evidenced by their weak spatial connection to massive galaxies.
The first, mentioned by Melnick~(\cite{Melnick87}) and clearly formulated
by Brinks~(\cite{Brinks90}), incorporates the tidal action
from nearby low-mass galaxies. It was tested in the VLA search for HI-rich
companions of 19 nearby  (V $<$ 2500 \kms) ``isolated''
\ion{H}{ii}-galaxies (Taylor et al.~\cite{Taylor95}; see also
Taylor~\cite{Taylor97}).
Surprising, 10 of the 19 target galaxies  revealed low-mass
HI companions, most of which were identified later with faint galaxies.
However, the statistics are still rather sparse.
Some additional indications of the potentially important role of low-mass
galaxies appeared after the deep CCD imaging of the large BCG sample by
Lipovetsky et al. (\cite{Lipovet}).

Another model of an external trigger for SF bursts in BCGs was suggested by
 \"Ostlin et al.~(\cite{Ostlin99}) and Kunth \& \"Ostlin~(\cite{Kunth2000}).
Based on morphology and gas kinematics data for the small
sample of luminous BCGs, they consider that mergers of low-mass galaxies are
the most important factor affecting the group of BCGs. Similar results were
reported by Sung~(\cite{Sung00}), based on studies of a sample of
115 BCG/\ion{H}{ii}-galaxies.

Thus, during the last 5--7 years, an increasing amount of evidence has
appeared which imply the importance of gravitational interaction of low-mass
galaxies with current SF bursts, in particular with galaxies of
comparable or lower mass.
At the same time, neither new simulation has demonstrated the
potential of an internal spontaneous onset of SF bursts in gas-rich galaxies,
nor do any concrete models indicate the physical conditions which are
necessary for this mechanism to apply.

Therefore the relative role of external and internal
trigger
mechanisms of SF in BCG/\ion{H}{ii}-galaxies seems far from to being settled.
It is high time to pose the question: what are the main trigger
mechanisms of SF burst in BCGs? This is important both for the general
understanding of the BCG phenomenon and the cosmological evolution of BCGs.

In this paper, we use the well-selected BCG sample of 86 galaxies in the zone
of the Second Byurakan Survey  to get the quantitative estimate of the
relative frequency of gravitational interaction as a probable trigger of SF
bursts in BCG progenitors.
Based mainly on UZC (Falco et al.~\cite{Falco99}) and
the NED\footnote{http://nedwww.ipac.caltech.edu/ -- the
NASA-IPAC Extragalactic Database},
we perform an extensive search for nearby disturbing galaxies, both more
massive and those of comparable or lower luminosity than that of the studied
BCGs.
For BCGs without such probable disturbers in UZC and NED and in recent
publications on
galaxy redshifts, we have measured the radial velocities of some candidate
disturbers with the SAO 6\,m telescope.

The paper is organized as follows. In section 2 we briefly review tidal
trigger of SF, discuss in more detail the most efficient examples, and
describe the criteria we apply to assign a nearby galaxy as a disturbing
neighbour.
In section 3 we describe our BCG sample, while in section 4 we present the
results of the ``companion'' search  and explain the sources we have used.
A brief analysis of the properties of galaxies, associated with BCGs, the
discussion of results obtained and our conclusions are presented in section 5.
The results of spectroscopy of 3 BCGs and 24 candidate companions in the
vicinity of 15 studied BCGs are presented in the Appendix.

\section{Tidal trigger}

{\it Enhanced} SF activity in large gas-rich galaxies is
stimulated by various types of galaxy interactions, including:
a) mergers of close mass galaxies (e.g., Larson \& Tinsley~\cite{Larson78};
Joseph et al.~\cite{Joseph84});
b) sinking dwarf companions (e.g., Mihos \& Hernquist~\cite{Mihos94};
Rudnick et al.~\cite{Rudnick2000});
c) strong tidals (e.g., Bernloehr~\cite{Bernloehr93}; Keel~\cite{Keel93});
d) weak tidals (e.g., Icke~\cite{Icke85}; Reshetnikov \&
Combes~\cite{Reshet97}; Rudnick et al.~\cite{Rudnick2000}).
In principle, all these mechanisms could also work for gas-rich low-mass
galaxies. The relative importance of these external triggers,
as well as their role in comparison to some internal triggers,
seemingly should depend on the type of environment.

While a detailed understanding of trigger mechanisms of enhanced SF due
to the tidal action of a nearby galaxy on gas-rich dwarfs is still lacking,
several schemes were suggested to model the effect of galaxy interaction,
which can be applied to low-mass galaxies.
The method of Noguchi~(\cite{Noguchi88}) is based on the generation of
a central bar
which in turn disturbs gas clouds and causes them to sink towards the center
of galaxy, inducing a central SF burst. Another mechanism, suggested by Olson
\& Kwan~(\cite{Olson90}), works via a large tidal increase in the inelastic
collision rate of individual gas clouds, and their merging and collapse. It
is basically a stochastic process. Both these mechanisms require strong
enough tidal disturbance, possible when the fly-by of equal mass galaxies
takes place at the pericenter distance of about 8--10 disk scale-lengths.

The mechanism suggested by Icke~(\cite{Icke85}) involves the generation of
shocks in the outer parts of the HI disk, with the subsequent dissipation of
their kinetic energy and the loss of dynamical stability of the gas disk.
It takes a
moderate tidal force, and under similar conditions effectively acts from
a distance 2--3 times larger than the two former models.
Also, the cross-section of the Icke mechanism is several times larger.

This circumstance favors the Icke mechanism as the most efficient in the
stimulation of abnormal SF activity. Some indirect arguments to support
the importance of this trigger mechanism are the traces of recent
enhanced SF  in spiral galaxies with lopsidedness and bent outer parts. The
latter are considered to be evidence of minor tidal effects (e.g. Rudnick et
al.~\cite{Rudnick2000}; Reshetnikov \& Combes~\cite{Reshet97}).
However, lopsidedness can be caused by minor mergers as well (Rudnick et al.
~\cite{Rudnick2000}). Evidently, the two former, or similar, mechanisms also
can work to trigger BCG progenitors. However, their role is probably less
important.

To illustrate quantitavely the parameters of the Icke mechanism and the
pericenter distance at which tidals become effective at generating shocks, we
give below the simple formulae taken from the paper by Icke (\cite{Icke85}).

The Icke mechanism is based on tidal acceleration
of gas layers to supersonic velocities and the subsequent dissipation
of shock waves.  A characteristic value for the speed gained by
gas due to the tidal acceleration by an ``intruder'' (the object which
exerts the tidal force; the name does not necessarily imply a close approach)
over a time span corresponding to one revolution (a characteristic time in
the galaxy--``victim'' system) is given by the formula:

$\delta v$ $\approx 4GMR\cdot$$r^\mathrm{-3} \cdot$$2\pi R/v$ = $8\pi\cdot
v(R/r)^\mathrm{3}$,

\noindent
where $M$ is the mass of the galaxies, assumed here to be equal, $R$ - is the
radius of the external part of the galaxy--victim disturbed by the
tidal interaction, $r$ is the distance between the galaxy centers, and $v$
is the circular speed at the radius $R$, assumed to be equal to $v$ =
$(GM/R)^\mathrm{1/2}$.  From the condition that the shock can occur, that is,
$\delta v$ $> s_\mathrm{0}$ ($s_\mathrm{0}$ is ``sound'' speed), one gets
the condition
for the pericenter distance $p_\mathrm{0}$, at which the tidal interaction
will lead to shock generation:
$p_\mathrm{0} \approx$$R\cdot$($8\pi\cdot$$v/s_\mathrm{0})^\mathrm{1/3}$.

For the general case of colliding galaxies of unequal masses, with the
mass ratio $\mu$ =
M(intruder)/M(victim)
the latter relation changes to:
$p_\mathrm{0} \approx$$R\cdot$($8\pi\cdot\mu$ $v/s_\mathrm{0})^\mathrm{1/3}$.

To illustrate the range of the expected values of $p_\mathrm{0}$ for several
representative cases of the mass ratio, we calculate this according to the
formulae above for two low-mass "disk" galaxies with typical of BCG
parameters.
We accept here $s_\mathrm{0}$=10 \kms\ as the characteristic velocity
dispersion in the interstellar gas.  One case corresponds to a galaxy with
M(victim)$_\mathrm{tot}$ = 10$^\mathrm{10}$ M$_{\odot}$;  $ v/s_\mathrm{0}$=7;
R(victim)=10 kpc
and  another -- to a galaxy with
M(victim)$_\mathrm{tot}$ = 10$^\mathrm{9}$ M$_{\odot}$;  $ v/s_\mathrm{0}$=3;
R(victim)=5 kpc.
R(victim) is taken here as the size of the outer HI-disk, corresponding to
the characteristic mass density of $\sim$0.5 M$_{\odot}$/kpc$^\mathrm{2}$.
We notice that according to observations, BCGs, as well as other dwarfs,
are dominated dynamically by Dark Matter halos, and their baryonic mass is
several times lower than their total mass. So a DM halo plays the same role as
a spherical stellar halo in the original Icke model.
The respective values of $p_\mathrm{0}$ for $\mu$=200, 10, 1 and 0.1 are
shown in  Table~\ref{Tab_1}.
The parameters above correspond to the typical ones of BCGs, as described,
e.g., by Taylor et al.~(\cite{Taylor95}). According to the same authors, BCGs
with such total masses can have a wide range of blue luminosities,
corresponding to M$_\mathrm{B} = $ --12.7 to --17.0.

\begin{table}[h]
\centering{
\caption{Characteristic threshold pericenter distance}
\label{Tab_1}
\begin{tabular}{rrrrrr} \\ \hline
 \MC{1}{c}{M(vict)$_\mathrm{tot}$} &
 \MC{2}{c}{10$^\mathrm{10}$M$_{\odot}$} &
 \MC{1}{c}{} &
 \MC{2}{c}{10$^\mathrm{9}$M$_{\odot}$} \\  \hline

M(intrud)                    &$\mu$  & $p_{0}$      &M(intrud)   &$\mu$  & $p_{0}$      \\
(M$_{\odot}$)                &       & (kpc)        & (M$_{\odot}$)                  &        & (kpc) \\   \hline
2$\cdot$10$^\mathrm{12}$     &200.0  & 327          & 2$\cdot$10$^\mathrm{11}$       &  200.0 & 123   \\
10$^\mathrm{11}$             & 10.0  & 121          &  10$^\mathrm{10}$              &  10.0 &  45   \\
10$^\mathrm{10}$             &  1.0  & 56           &  10$^\mathrm{9}$~              &  1.0  &  21   \\
10$^\mathrm{9}$~             &  0.1  & 26           &  10$^\mathrm{8}$~              &  0.1  &  10    \\   \hline
\end{tabular}
}
\end{table}

One important note related to the problem of the observational search for
potential ``intruders'' around the target BCGs is connected to the expected
time delay between the passage of the pericenter and the beginning of SF burst.
For example, for V$_\mathrm{rot}$= 70 \kms\ and $R$=10 kpc, the characteristic
time of development of shocks is comparable (Icke \cite{Icke85})
to $T$=2$\pi$ $R/V_\mathrm{rot}$ = 8$\times$10$^\mathrm{8}$ yr.
In general, checking any nearby galaxy as a potential ``intruder'' of the
target BCG, we need to account for the relative tangential velocities of the
two considered galaxies $\delta$V, in our case up to 300--400 \kms.
Thus, e.g., for $\delta$t = $T$/2 for the disk above, and $\delta$V = 200
\kms, by the time of a well-developed SF episode, the two galaxies
can increase their projected separation from their pericenter position by
90 kpc.
Hence, due to the time delay between the maximum tidal action and
its consequence as a SF burst, one can expect that
even relatively low-mass neighbours can be the real disturbing galaxies if
they are found at distances of less than $\sim$100--150 kpc. The delays in
the beginning of a SF burst in the more massive component in pairs of
interacting spirals with highly different masses were noticed by
Bernloehr~(\cite{Bernloehr93}). These delays
range up to several hundred Myr, which in general is compatible with the
numbers given above.

Sufficiently detailed observational data for BCG companions to be
used to test the role of relatively weak tidals are not abundant. The most
suitable are seemingly the results of the search for low-mass HI companions
near \ion{H}{ii}-galaxies by Taylor et al.~(\cite{Taylor93, Taylor95}).
The projected distances in the systems with detected HI companions are
consistent with the estimates of the threshold pericenter distances with
input observational parameters, if again, some reasonable delay in the
ignition of SF bursts is taken into account.

We also  note that to directly detect
shocks in HI gas of BCGs triggered by an interaction is not simple.
According to the model, they dissipate to initiate gas instability.
Therefore, by the time of the developed SF burst, the signs of shocks
can be completely erased. Therefore the search for such disturbed velocity
fields should be directed to the systems in the earlier stages of collision,
in which an SF burst has not yet occured. Probably gas-rich galaxy pairs with
a large mass difference, such as those studied by Bernloehr
(\cite{Bernloehr93}), can be suitable for this task.

\section{BCG sample}

The BCG sample for this study is compiled from the BCG sample in the zone
of the Second Byurakan Survey (SBS)
(R.A.=7$^\mathrm{h}$40$^\mathrm{m}$--17$^\mathrm{h}$20$^\mathrm{m}$,
Dec=49$^{\circ}$---61$^{\circ}$).
The latter is partly described by Izotov
et al.~(\cite{Izotov93}), Thuan et al.~(\cite{Thuan94}),
Pustilnik et al.~(\cite{Pustilnik95}). The
same BCG sample was in particular the subject of an
HI-study by Thuan et al.~(\cite{Thuan99}). Their subsample of 88 BCGs is
reduced slightly here to match the strict criterion of the strength of the
[O{\sc iii}]-line.

Our sample  includes 86 BCGs with a sufficiently strong [O{\sc iii}]-line
(EW([O{\sc iii}]$_\mathrm{\lambda5007}$) $>$ 45 \AA) and V$_\mathrm{hel}<$
6000 \kms.
The first criterion corresponds to the line strength at which the loss of
candiadte emission-line galaxies (ELGs) from objective prism plates during
the primary selection
is moderate, while the second criterion is to limit the volume where search
for non-massive neighbours can be efficiently carried out.
While the question of completeness analysis of SBS BCG sample is beyond
the scope of this paper, it is useful to describe the bias inherent to the
BCG sample. As demonstrated by Salzer~(\cite{Salzer89}), the selection
in emission-line surveys is determined by the total flux in line+continuum.
Thus, for apparent magnitudes close to the threshold of photographic plates,
only emission lines with large enough equivalent widths (EWs) will be
detected. From the carrying out
of such selection on the SBS plates and the preliminary comparison
in the small overlapping region of SBS detected BCGs with BCGs detected in the
Hamburg/SAO survey (Ugryumov et al.~\cite{Ugryumov99}), for the limiting
m$_\mathrm{B}$=18, which is near the completness limit of the SBS sample
(Pustilnik et al.
\cite{Pustilnik95}), galaxies having \ion{H}{ii}-regions with
EW([O{\sc iii}]$_{\lambda5007}$)
$>$(45--50)\AA\ are selected with modest loss, and thus can be considered as
representative of the whole population of galaxies with that level of SF
activity.
This level corresponds, for typical extragalactic \ion{H}{ii}-regions,
to EW(H$\beta$)$>$(10--20)\AA, which in turn in the model of instantaneous
SF burst corresponds to the burst age $<$(7--10) Myr (Leitherer et al.
\cite{Starburst99}).  Therefore, the main bias introduced by the applied
criteria
to the entire BCG/\ion{H}{ii}-galaxy population is the selection of relatively
young SF bursts, which can more easily be found in objective prism spectral
plates.
Some characteristics of the sample are: the range of
apparent blue magnitudes
B = 12\fm5 $\div$ 18\fm5, the range of absolute blue magnitudes
M$_B = -11\fm0 \div -19\fm4$. For many of these galaxies, the accuracy of
radial velocities was significantly improved after HI-line measurments by
Thuan et al. (\cite{Thuan99}) and additional optical spectroscopy
(Izotov et al.~\cite{Izotov94}; Thuan et al.~\cite{Thuan95}).
To illustrate the properties of this BCG sample, we show in
Fig.~\ref{fig:Fig_ab} some of its parameter
distributions (m$_\mathrm{B}$, M$_\mathrm{B}$, V$_\mathrm{hel}$).

\begin{figure}[hbtp]
    \psfig{figure=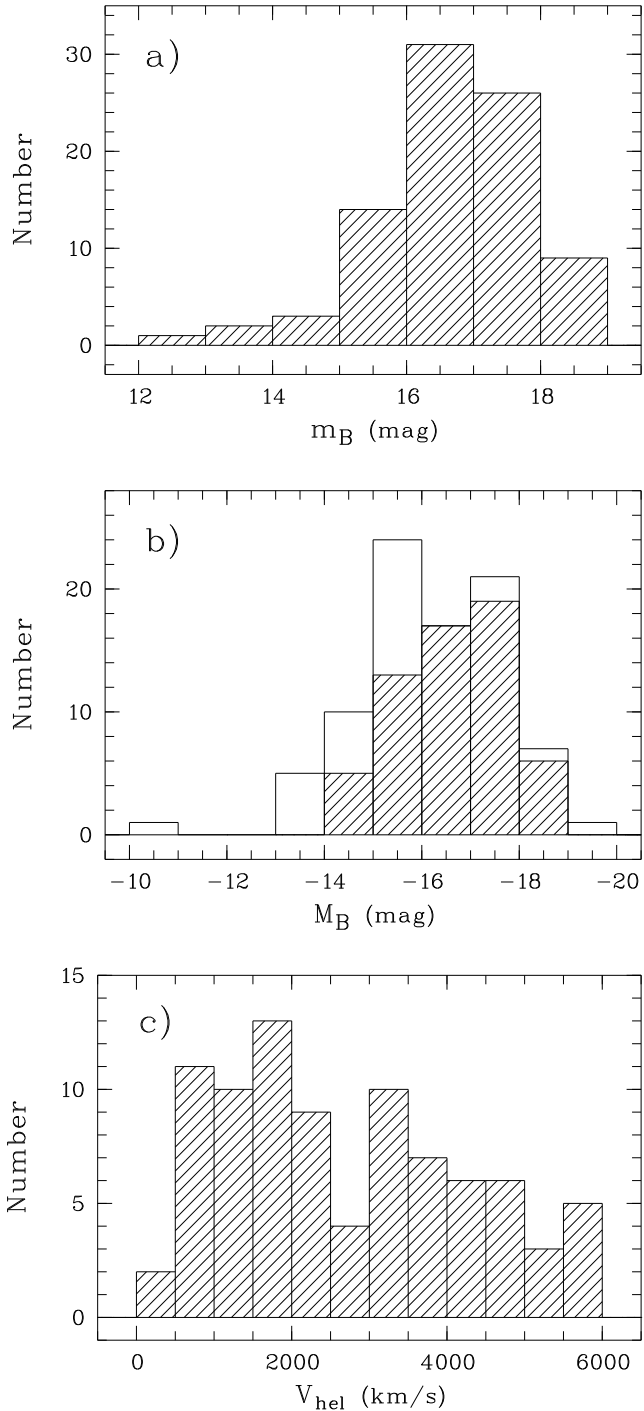,angle=0,width=8.5cm,bbllx=40pt,bblly=250pt,bburx=300pt,bbury=830pt}
    \caption{Distributions of the BCG sample
     on m$_\mathrm{B}$ (a), M$_\mathrm{B}$ (b) and V$_\mathrm{hel}$ (c).
     The hatched and unhatched parts of the M$_\mathrm{B}$-distribution
     separate BCGs from the general field group
     (60 galaxies) and the Local Supercluster volume (26 galaxies) respecively.
     See subsection~\ref{subsection:Effect} for details .
     }
    \label{fig:Fig_ab}
\end{figure}

One can see that the sample is more or less typical, with some excess
of the less luminous BCGs in comparison to e.g., the BCG sample in the same
sky region studied by Pustilnik et al.~(\cite{Pustilnik95}). This is connected
to the different velocity boundaries adopted for the latter paper (2,000
\kms\ $<$ V$_\mathrm{hel}$ $<$ 10,000 \kms) and ours
(V$_\mathrm{hel}$ $<$ 6,000 \kms).
This sample includes a significant number of BCGs located inside the Local
Supercluster that could bias the conclusions relative to the situation in
the general field.
This question will be addressed in the discussion section.
B-magnitudes of many BCGs in the considered sample are significantly
corrected based on the results of CCD-photometry (Lipovetsky et
al.~\cite{Lipovet}; Kniazev et al.~\cite{Kniaz_Loi}).

The list of the studied sample of 86 BCGs is
presented in Table~\ref{Tab_2}, containing the following information: \\
 {\it column 1:} The object's IAU-type name. For those without an alternative
name the prefix SBS is applied in the last column. \\
 {\it column 2:} Right ascension (R.A.) for equinox B1950. The coordinates
  of BCGs in this table sometimes differ from those presented in NED, but
  are the most up-to-date, according to our own checks on the Digitized Sky
  Survey (DSS). \\
 {\it column 3:} Declination for equinox B1950.  \\
 {\it column 4:} Heliocentric velocity in \kms. For more than 60
   BCGs the r.m.s. uncertainty is less than 30 \kms (mainly from HI data).
   For seven BCGs it can reach $\sim$100 \kms, and for the remaining
   galaxies it is $\sim$50--60 \kms. \\
 {\it column 5:} Reference to the source of velocity accepted, coded by
1 -- Thuan et al.~(\cite{Thuan99}):
2 -- NED;
3 -- Pustilnik et al.~(\cite{Pustilnik95});
4 -- new measurments with the SAO 6\,m telescope (this paper, Appendix);
5 -- UZC: Falco et al.~(\cite{Falco99});
6 -- Carrasco et al.~(\cite{Carrasco98}). Some of our BCG velocities in NED,
taken from HI-measurments by Thuan et al.~(\cite{Thuan99}), are not reliable,
since they ignore the information on real or possible confusion. We used
instead for these galaxies the velocities from Pustilnik et
al.~(\cite{Pustilnik95}).\\
 {\it column 6:} Distance in Mpc, accounting for Virgo infall correction from
Kraan-Korteweg (1986) with V$_\mathrm{infall}$ = 220~\kms. For one nearby
galaxy (V$_\mathrm{hel}$ = 289 \kms),
marked by an asterisk, the distance is accepted from Georgiev et
al.~(\cite{Georgiev97}). For another similar galaxy (Mkn~178,
V$_\mathrm{hel}$ = 249 \kms), the distance is
accepted to be equal to that of its neighbours on the sky with close
radial velocities and having photometrical determinations. \\
 {\it column 7:} Total B-magnitude from the unpublished CCD data
(Lipovetsky et al.~\cite{Lipovet}; Kniazev et al.~\cite{Kniaz_Loi}).
They are marked by a ``plus" before the value in the corresponding column.
For several galaxies, B-magnitudes are based on the APM database values
(Irwin \cite{Irwin98}). They were recalculated to the standard
CCD B-magnitudes using the relation between B$_\mathrm{CCD}$ and
B$_\mathrm{APM}$ derived
on more than 100 galaxies, as described by Kniazev et al. (\cite{Kniaz_HSS}).
Its r.m.s. uncertainty is found to be  0\fm45.
The B-magnitudes obtained in this way are marked by an ``asterisk"
before the value.
For a couple of objects, photometrical B-magnitudes are taken from NED.
They are marked by a ``n'' before the value. \\
 {\it column 8:} Absolute B-magnitude, calculated from the apparent
magnitude and the distance in column 5, with no correction for the Galaxy
extinction, since it is small for considered sky region. \\
 {\it column 9:}  A flag to distinguish the galaxies belonging to the
groups of the two types of environment (LS - Local Supercluster, GF -
general field). See subsection  (\ref{subsection:Effect}) for details. \\
 {\it column 10:}  One or more alternative names,
according to the information from NED.

\section{Search for neighbours and results}

\subsection{Search for neighbouring galaxies}

We employed the following sequence for the search for neighbours of the sample
BCGs. As a first pass we checked possible nearby galaxies in UZC, using the
criterion $| \Delta$V$| <$ 500 \kms\ and d$_\mathrm{proj} <$ 700 kpc. For the
second
pass, we used NED in order to check whether some fainter galaxies are present
in the BCG environment which could exert a stronger tidal action than the
candidates found (if any) from UZC. The same criterion was applied as for the
first pass, and then the candidates were additionally checked for
correspondance of their projected distances to those for potential intruders,
illustrated in Table 1. We also used some unpublished results on the
redshifts of SBS galaxies, as well as our own results of dedicated
spectroscopy of potential candidate companions.
Since the galaxies, picked up in NED, present a
mixture of the content of many catalogs as well as some small groups found
in surveys of varying powers, it is difficult to address the question
of completeness for the group of neighbouring galaxies found in the vicinity
of BCGs. Therefore we can only emphasize that our estimates of neighbouring
galaxies found this way are the {\it lower limits} of real numbers.

\subsection{Results and types of neighbours}

The results of the search for probable disturbing/interacting galaxies with
the sufficiently strong tidal action on the sample BCGs are summarized in
Table~\ref{Tab_3}.
We present the candidate disturbing galaxy  which exerts the
strongest tidal action on the target BCG. If there are other galaxies with
comparable effect, we indicate them in the Notes on individual BCGs after the
table.
Table~\ref{Tab_3} contains the following data: \\
 {\it column 1:} BCG IAU-type name. The same as in Table~\ref{Tab_2}. \\
 {\it column 2:} The name (or one of the best known names, such as NGC one)
 of the suggested disturbing galaxy. Prefix ``S'' in the name  means the
object is from SBS. Prefix ``A'' means a new object (anonymous), found
 as an associated galaxy in the spectroscopy data from the 6\,m telescope.
 In the case of a BCG supposed to be a merger, its name is repeated in this
 column. If neither merger morphology nor disturbing galaxy is identified,
this field is left blank. \\
 {\it column 3 and 4:} R.A. and Declination of the disturbing galaxy for the
equinox B1950. Not given for a suggested merger, or if no disturbing
galaxy is identified.\\
 {\it column 5:} Heliocentric velocity of BCG, in \kms.  Same as in
Table~\ref{Tab_2}.\\
 {\it column 6:} Heliocentric velocity of the disturbing galaxy, in
km~s$^{-1}$. When the velocity of the disturbing galaxy, originally found in
UZC, was known in NED with a better precision, we cite the NED value. \\
 {\it column 7:} The source of the value in the previous column, same coding
as in the description for Table~\ref{Tab_2}. \\
 {\it column 8:} B-magnitude of the disturbing galaxy. Most are taken
from NED. The rest are either recalculated from the APM B-magnitudes through
calibration with the CCD-based B-magnitudes, similar to those in
Table~\ref{Tab_2}, as described by Kniazev et al.~(\cite{Kniaz_HSS}),
and marked by ``*'' before the value, or are derived from the same
CCD-frames on which target BCGs were measured. These are marked
by ``+''.   \\
 {\it column 9:} Absolute B-magnitude of the disturbing galaxy, calculated
  with the same distance modulus as for the neighbouring BCG. \\
 {\it column 10:} Difference of the absolute magnitudes of BCG and its
  disturbing  galaxy $\Delta$M$_\mathrm{B}$ = M$_\mathrm{B}$(partner) --
  M$_\mathrm{B}$(BCG).
  A negative value means that the disturbing galaxy is brighter. \\
 {\it column 11:} The projected distance between the BCG and the disturbing
   galaxy, in kpc.\\
 {\it column 12:} Suggested trigger classification: p -- parent galaxy,
  when the disturbing galaxy is significantly brighter than the target
  BCG;  b -- binary system, in which the target BCG and its disturbing galaxy
  are dynamically comparable; f -- fainter companion,  when
  the disturbing galaxy is significantly less massive than the target BCG;
  m and m? -- merger morphology with the various degrees of confidence. The
  latter classification was performed through the comparison of the morphology
  of candidate BCGs on the CCD frames from Lipovetsky et al.~(\cite{Lipovet})
  and Kniazev et al.~(\cite{Kniaz_Loi}), or on the DSS-2 images, with the
  galaxies shown by Keel \& Wu~(\cite{Keel95}). \\

\begin{figure}[hbtp]
    \psfig{figure=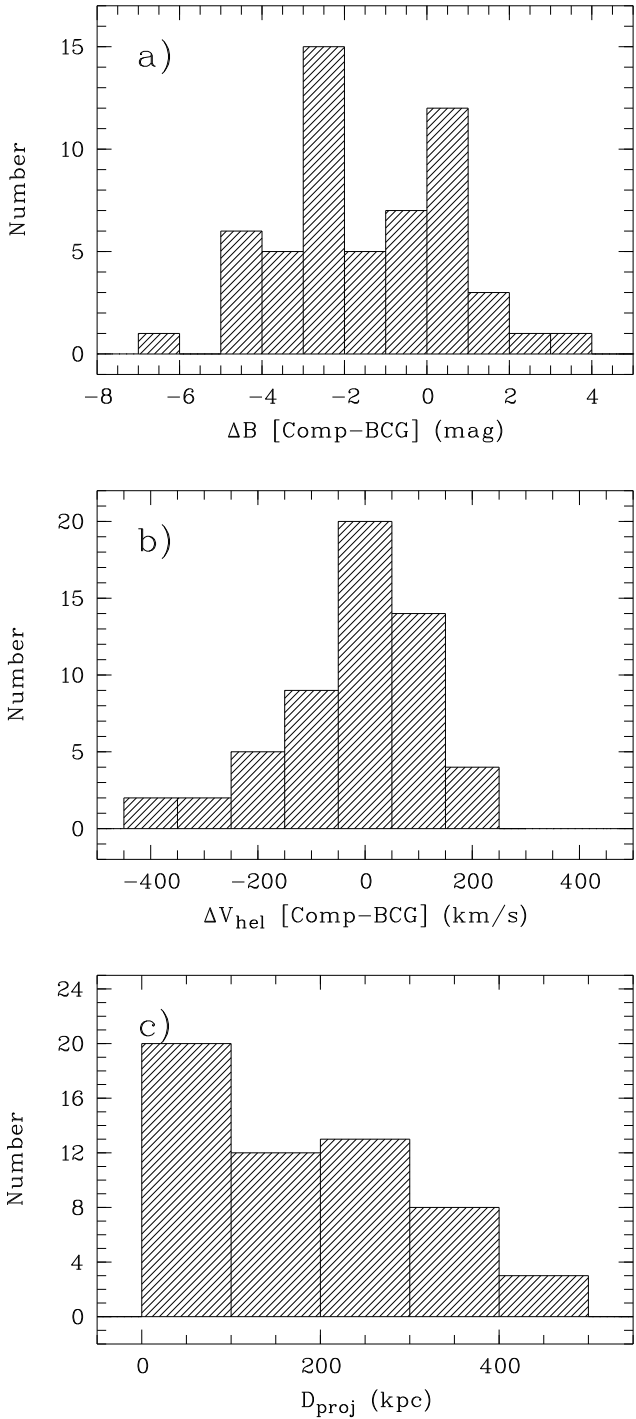,angle=0,width=8.5cm,bbllx=40pt,bblly=250pt,bburx=300pt,bbury=830pt}
    \caption{(a) Distribution of blue absolute magnitude diffirences of BCGs
     and their disturbing galaxies;
     (b) Distribution of the relative radial velocities of BCGs and their
     disturbing galaxies;
     (c) Distribution of the projected distances between BCGs and their
     partner galaxies}
    \label{fig:FigDist}
\end{figure}

From the data in Table~\ref{Tab_3}, one can see that the most
representative group of 33 BCGs in this sample
($\sim$38$\pm$7\%)\footnote{ Here and in other presented fractions
the errors are given based on Poisson statistics of numbers of BCGs
involved in each group.}
have close enough
brighter galaxies (conditionally classified as parent) with
$\Delta$M$_\mathrm{B}$ in the range --1.5$^\mathrm{m}$ to --6$^\mathrm{m}$.
23 BCGs from this sample ($\sim$27$\pm$6\%) are found to have disturbing
galaxies of comparable brightness, or even significantly fainter. Of the
remaining BCGs, which have no evident `neighbours', 14 BCGs
($\sim$16$\pm$4\%) show more or less evident merger morphology. Only 16
BCGs ($\sim$18.5$\pm$5\%) have no clear indication
of interaction with other galaxy. However,
for many of these BCGs there are faint galaxies in their vicinity with still
unknown redshifts.
So the fraction of `non-interacting' BCGs presented here should be rather
treated as an upper limit.

Thus, our results indicate that gravitational interaction between gas-rich BCG
progenitors and  various mass galaxies in their environment may play a key
role in the ignition of a SF burst. In particular, the conclusions by Taylor
et al.~(\cite{Taylor95}, \cite{Taylor97}) on the important role of
low-mass companions in triggering SF in ``isolated'' \ion{H}{ii}-galaxies are
consistent with our data. Similar conclusions are drawn by Noeske et
al.~(\cite{Noeske2001}), based on the environment study of BCGs, mainly
from the UM survey.
The obtained results do not exclude that some internal mechanisms, which
can trigger SF bursts in BCGs, probably work in the minority of them under
certain conditions.
However, in the majority of BCGs, the external trigger should be more likely,
connected to interactions with other galaxies.

\section{Discussion and conclusions}

\subsection{Companion properties}
\label{subsection:prop}

It is interesting to examine the properties of the found disturbing
galaxies and to describe in more detail the relative fractions of various
combinations of BCGs with their partner galaxies.

In Fig.~\ref{fig:FigDist} the distributions of blue absolute magnitude
differences, the relative projected distances and the relative radial
velocities are shown.

One can see that the general distribution of the projected distances
for suggested disturbing galaxies spreads upto $\sim$400 kpc, which is
consistent with conclusions by Zaritsky et al.~(\cite{Zaritsky94};
\cite{Zaritsky97}) on the distribution of companions around massive spirals.
However, if we limit the group of disturbing galaxies by ``dwarfs''
(M$_\mathrm{B} \geq -$18\fm5), then the distribution of the projected
distances is much narrower, with the median D$_\mathrm{proj} \sim$ 90 kpc.
The distribution of the relative radial velocities of the partner galaxy and
the BCG
peaks at $\Delta V$=0, with $\sim$93\% of this value in the range
$\pm$250 \kms. The latter value is typical for these studies,
considering possible physical neighbours of field galaxies.

In Fig.~\ref{fig:FigDist} the full range of total magnitude differences is
shown for BCGs with identified disturbing partners. Its asymmetric (relative
to zero) and bimodal
appearance at least partly is caused by selection effects. It is more
difficult to measure the redshifts for the fainter candidate neighbours.
However, even if all remaining BCGs without identified disturbing galaxies
would
appear to have fainter companions after a more careful search,
this distribution will be still bimodal, with the prominent peak at $\Delta
B$ around --2\fm5. Accounting for the mean brightening of BCGs due to a SF
burst of the order $\Delta B=$0\fm75--1\fm0 (e.g., Papaderos et al.
\cite{Papa96}; Lipovetsky et al.~\cite{Lipovet}),
the full interval of the mass ratios
of the BCG progenitor and its disturbing galaxy ranges from $\sim$1/400 to
$\sim$25. The peak in the distribution of $\Delta B$ indicates that
the most common brighter disturbing galaxies are more massive than the BCG
progenitors by a factor of 10--25.

\subsection{The problem of a control sample}

For confident conclusions on the results of statistical studies like this, it
is quite important to compare them to the results of
a similar study on a control sample. The objects of the latter should be
separated to have a complimentary property relative to that of the main
sample.
In our case, it should be a sample of gas-rich, low-mass galaxies with no SF
bursts. However, while there is no problem in obtaining a large, well-selected
BCG
sample in the considered volume, it is very difficult to form such a sample
of LSB dwarfs or Im galaxies.
We postpone the similar
study of control samples to the time when such samples will be available,
hopefully after SDSS (Sloan Digital Sky Survey)
data will be distributed.
For the purposes of illustration, we undertook the study of the
environments for the sample of LSB dwarfs by Pildis et al.~(\cite{Pildis97}).
We separated from their 110 LSB dwarfs, 56 objects with V$_\mathrm{hel}  <$
6,000 \kms\ and V$_\mathrm{Vir} >$2,000 km~s$^{-1}$, that is, with the
same criteria as for our general field (GF) BCG group (see next subsection).

\begin{figure}[hbtp]
    \psfig{figure=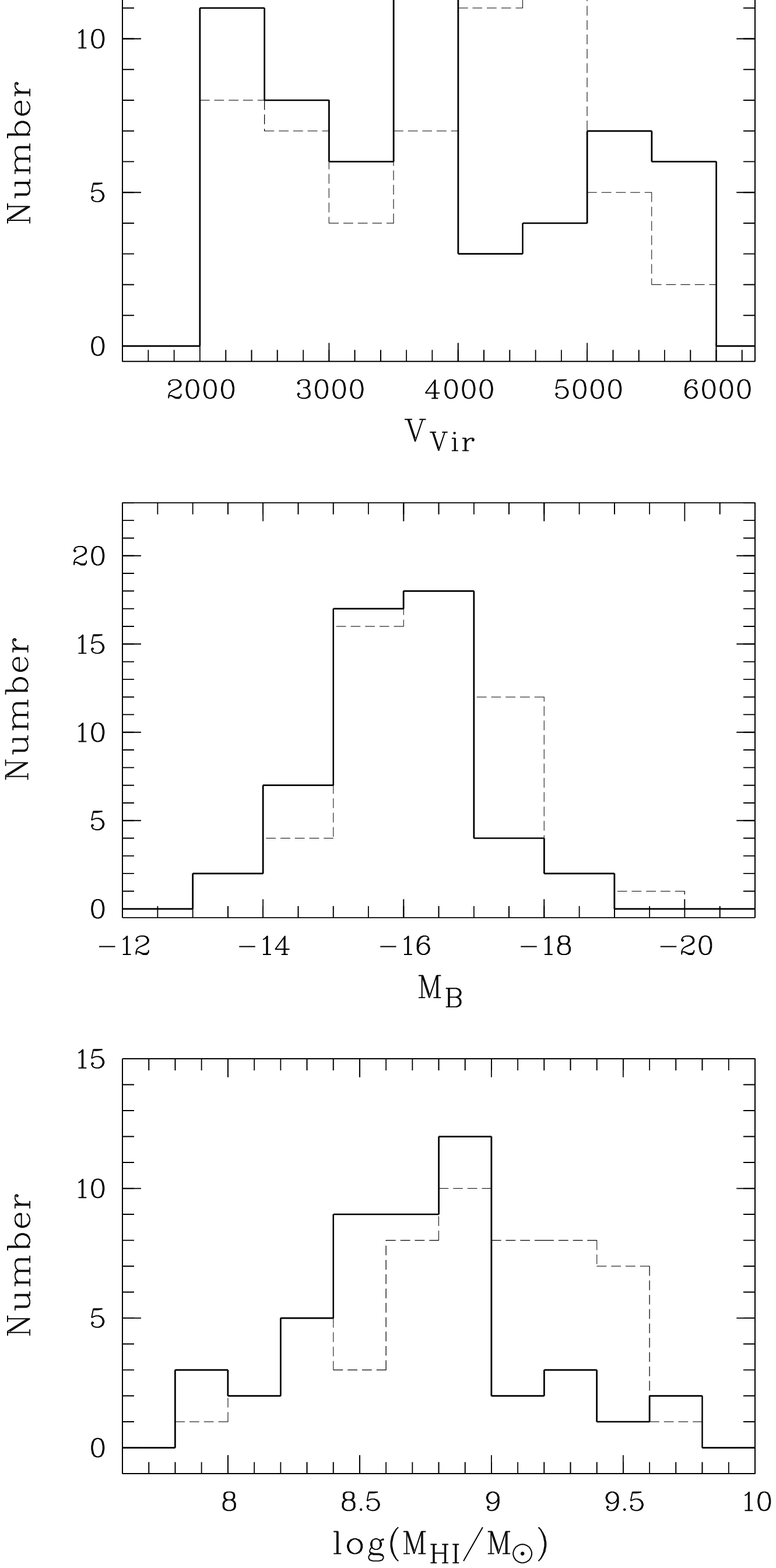,angle=0,width=8.5cm,bbllx=10pt,bblly=6pt,bburx=433pt,bbury=838pt}
    \caption{Distributions of two compared samples -- 49 BCGs from the
     general field (GF) region with M(HI) data (see \ref{subsection:Effect})
     (solid line) and 56 LSB
     galaxies from Pildis et al.~(\cite{Pildis97}) (dashed line), selected in
     the same velocity range. \newline
     {\it Upper panel:} distribution of radial velocity;
     {\it Middle panel:} distribution of absolute $B$-band magnitude;
     {\it Lower panel:} distribution of M(HI) (in solar masses).
     }
    \label{fig:LSBaBCG}
\end{figure}

Before comparing this group of LSB galaxies with BCGs for the presence
of disturbing neighbours, it is reasonable to check how similary they are
distributed in space (that is, in radial velocity), and how similar are their
ranges of global parameters, like luminosity and gas mass.
In Fig.~\ref{fig:LSBaBCG} we show distributions of V$_\mathrm{r}$,
L$_\mathrm{B}$,  and
HI mass for both samples. We note that V$_\mathrm{r}$ and
L$_\mathrm{B}$ distributions for BCGs and LSB galaxies are similar.
The null hypothesis on the same distribution of both samples can
not be rejected, according to the Kolmogorov-Smirnov $D$ statistic, even on
the significance level 0.37 and 0.31, respectively. This makes the comparison
of distances to the disturbing galaxy secure for the two samples .

For M(HI), despite the closeness in the total range of this parameter for
both samples, their distributions have larger differences, which is also
reflected in the Kolmogorov-Smirnov $D$ statistic (significance level for
rejection of the null
hypothesis is 0.03). This difference, however, is only of the
general interest. The amount of HI in the considered galaxy samples can be
the result of the difference of their distances to the disturbing
neighbours, but can not  be the reason for this difference.

To compare the distributions of LSB galaxies and BCGs according to the
distance to
the most disturbing galaxy, we proceeded the same way as for BCGs, but
have checked potential disturbers at the projected distances upto 3 Mpc.
First, we checked potential neighbours in UZC (Falco et
al.~\cite{Falco99}), then looked at possible fainter neighbours
in NED. Finally, we selected the most tidally disturbing galaxies, assuming
mass to be proportional to L$_\mathrm{B}$, and the tidal force proportional
to the cubed inverse projected distance $D^\mathrm{-3}$.

The search for the strongest disturbing galaxies for this LSB subsample have
been made in UZC and NED, taking the limiting difference in radial
velocity $| \Delta V | \leq$ 500 \kms.
Its results are illustrated in Fig.~\ref{fig:Fig_LSB}
along with the results of a similar search for BCGs from the GF group.
For BCGs from Table~\ref{Tab_3} with no identified real tidal disturbers,
the strongest disturbers are found to be at the distances 600 -- 2000
kpc, which leads to the long tail. BCGs with a merger morphology were excluded
from the comparison with LSB galaxies.

Summarizing, we draw the following conclusion. BCGs/\ion{H}{ii}-galaxies
significantly more often have the strongest disturbing neighbours at a
distance of tens to a few hundred kpc than LSB dwarfs have.
E.g., $\sim$76\% of BCGs have such neighbours at
D$_\mathrm{proj} <$ 400 kpc
while for LSB dwarfs this fraction is only $\sim$46\%.
The median and mean values of D$_\mathrm{proj}$ -- 270 and 384 kpc for
BCGs, and 560 and 728 kpc, respectively for LSB dwarfs, differ by about factor
of two.
In addition,  $\sim$1/5 of BCGs show merger morphology.
If these 11 BCGs  from the general field are included in the first bin
D$_\mathrm{proj} <$ 100 kpc, the difference between
BCGs and LSB dwarf distributions becomes striking.

Bothun et al. (\cite{Bothun93}) made a comparison of the local
environment of a large LSB disk galaxy sample  to that of HSB (high surface
brightness) disks, and arrived a similar conclusion on the significant
statistical deficit of galaxies from the CfA redshift survey within a
projected radius of 0.5 Mpc from LSB galaxies.

\begin{figure}[hbtp]
    \psfig{figure=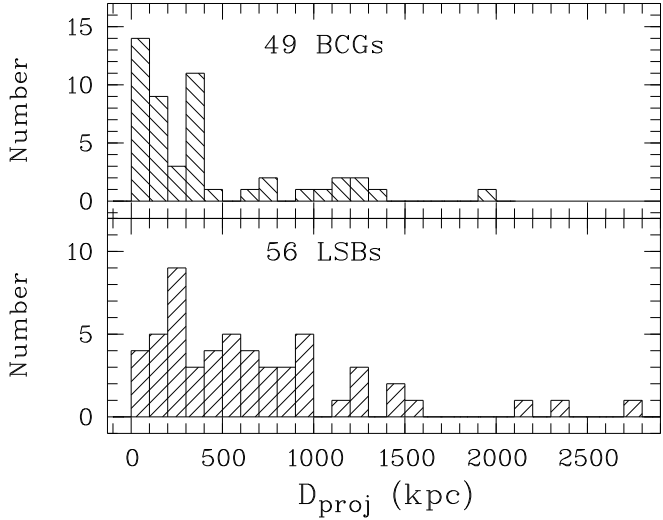,width=8.2cm,bbllx=45pt,bblly=50pt,bburx=555pt,bbury=655pt,angle=-90}
    \caption{Distribution of projected distances between the studied galaxies
     and their the most important disturbing neighbour, as found in NED. Upper
     panel: 49 BCGs from the general field group (see \ref{subsection:Effect}).
     Lower panel: 56 LSB galaxies from Pildis et al.~(\cite{Pildis97}).
     Median values are 270 kpc for BCGs and 560 kpc for LSBDs.
     }
    \label{fig:Fig_LSB}
\end{figure}

Other than this exercise on brighter neighbouring galaxies, we can cite the
results of the search for HI-rich low-mass companions for the small LSB
dwarf sample by Taylor~(\cite{Taylor97}).
It is shown that  the control sample of LSB dwarfs has a significantly lower
fraction of low-mass companions than the sample of BCG/\ion{H}{ii}-galaxies.
However, several low-mass HI-companions do exist
around LSB dwarfs, and they do not trigger SF bursts in these  galaxies. This
clearly indicates that the presence of a sufficiently close disturber cannot
be the only factor determining enhanced SF in gas-rich galaxies. One
plausible interpretation of this conclusion by Mihos et al. is mentioned in
subsection \ref{subsection:trigger}.
On the other hand, $\sim$40\%
of their BCG sample have no detected companions. While the effect of
incompletness probably affects this number, one can expect that some
fraction of ``isolated'' BCGs are not triggered by interactions with low-mass
companions.

Another sample, interesting for comparison, is the Local Volume sample (LV)
(e.g., Karachentsev \& Makarov~\cite{Kara99}), that is galaxies with radial
velocities relative to the centroid of Local Group V$_\mathrm{LG} <$
500 \kms.
Among 335 galaxies currently known in the LV (Makarov~\cite{Makarov00}), the
class of BCG (or post Star-burst) can be assigned  to the following objects:
NGC~1569, NGC~1560, SBS~1123+576, Mkn~178, VII~Zw~403, Arp~211, NGC~6789.
Of them SBS~1123+578 is probably tidally disturbed and Mkn~178 is a merger,
as shown in Table 3. The post Star-burst dwarf NGC~1569 interacts with
UGCA~92 at a distance of $\sim$40 kpc, or with a low-mass HI companion (Stil
\& Israel~\cite{Stil98}). The post Star-burst dwarf NGC~1560 is in the group
Maffei/IC~342 at a distance of $\sim$300 kpc from IC~342, which could
disturb it, accounting for the time elapsed since beginning of the SF burst.
For Arp~211, the situation is uncertain. Its photometrical distance of 2.8 Mpc
(Nordic Optical Telescope -- NOT data, Makarova et al.~\cite{Makarova98}),
after obtained HST data changed to 6.7 Mpc (Crone et al.~\cite{Crone01}).
Its probable strongest disturber, IC~3687, from the NOT photometry of 3
brightest blue stars is at about 3.0 Mpc. However, this estimate is
probably not reliable, since from the estimate on the brightest red star, the
distance is close to that of HST for Arp~211 (Makarova et
al.~\cite{Makarova98}), and Arp~211 can be disturbed by IC~3687.
The remaining BCGs, VII Zw~403 and NGC~6789 are probably well isolated.
So, summarizing, we
conclude that the data on a very limited number of star-bursting dwarfs in the
LV do not contradict the results of the study on 86 BCGs above.

\subsection{Effect of global environment}
\label{subsection:Effect}

One of the important points of this study is to determine if the results
obtained for the
discussed BCG sample are representative of the general situation with BCGs.
It is evident that some bias in the fraction of BCGs with the
galaxy-interaction-induced SF burst must be present in our sample, since a
significant number of our BCGs fall into the Local Supercluster, which is not
typical of the general field. In paricular, besides the environments of the
Virgo cluster, the galaxies of the UMa poor cluster (Tully et
al.~\cite{Tully96}) (with
the center at R.A.$\sim$12$^h$, Dec.$\sim$49$^{\circ}$, the angular diameter
of $\sim$13$^{\circ}$ and
the radial velocity range V$_\mathrm{hel}$ of 640 to 1150 \kms)
can appear among the BCGs of our sample.

The simplest way to check how much the effect of Local Supercluster
contributes to the various types of interactions is to divide our sample
by radial velocity. Roughly, the BCGs with the Virgo infall-corrected
velocity V$_\mathrm{Vir} <$ 2,000 \kms (D$_\mathrm{Vir} <$26.67 Mpc) are
considered constituents of the Local Supercluster volume, and all remaining
as the galaxies of general field. The respective total numbers for these
two groups are: 26 and 60 BCGs.

For the 26 BCGs within the ``borders'' of the Local Supercluster volume, we
have the following statistics. 14 of them ($\sim$54$\pm$14\%) have
significantly brighter disturbing neighbours ($\Delta B \leq$--1\fm5).
5 BCGs ($\sim$19$\pm$8\%) have neighbours of similar or lower
luminosity, and 3 ($\sim$11.5$\pm$6.5\%) have no neighbours but have
a merger morphology. 4 more BCGs  ($\sim$15.5$\pm$8\%) have neither
evident companions (or have possible
candidates with unknown velocities) nor show evidence of a recent
merger.

For the remaining 60 BCGs in the general field, only 19 ($\sim$31.5$\pm$7\%)
have significantly brighter disturbing neighbours. 18 BCGs (30$\pm$7\%)
have neighbours of similar or lower luminosity. 11 BCGs ($\sim$18.5$\pm$6\%)
have no disturbing
neighbours, but have more or less clear indication of merger morphologies.
The remaining 12 BCGs without evident companions or merging correspond to
20\%.

Comparing these two groups of BCGs from our sample, we see that the fraction
of BCGs probably triggered by interactions with other galaxies is similar
($\sim$84.5$\pm$18\% and 80$\pm$12\%, respectively) in the Local
Supercluster volume and outside of it.
The fractions of BCGs with various types of disturber and merger morphology
differ in the two groups. However, only the difference in the fractions of
BCGs with significantly brighter disturbing galaxies ($\sim$54$\pm$14\%
versus $\sim$31.5$\pm$7\%) is seemingly real (while it is still within
1.5$\sigma$ of the combined uncertainty).  This can quite naturally be
interpreted as an indication of the more important role of interactions of
BCGs with massive neighbours in the regions of higher galaxy density.
The differences in the
fractions of other categories (low-mass neighbours and mergers) are within
the statistical uncertainties accounting for the small numbers of BCGs
involved in the comparisons.

Thus, we conclude that in the general field, the role
of galaxy interactions to trigger SF in BCGs is as important as in the Local
Supercluster volume. $\sim$80\% of
all BCGs are probably either tidally disturbed or result from recent mergers.
At least $\sim$60\% of field BCGs experience a tidal disturbance from other
galaxies. Mergers and probable mergers constitute $\sim$1/5 of all BCGs.

Some fraction of the remaining $\sim$20\% of ``non-interacting'' BCGs can in
fact also be interacting. Several of them have massive galaxies at
projected distances of $\sim$500--600 kpc, and can be their distant
companions.
On the other hand, for several ``non-interacting'' BCGs, there are
close candidate faint companions without radial velocities.

\subsection{On the results of complimentary approach}

Telles \& Maddox~(\cite{Telles00}) made an attempt to address the problem
of faint companions of \ion{H}{ii}-galaxies through the 2D statistical
analysis of their environment using the APM galaxy catalog of the
southern sky.
They did not find an excess of faint galaxies around studied BCGs,
and concluded that ``tidal interactions cannot be the only factor that
triggers their burst of star formation''. We also do not generalize
interactions as the only trigger of SF bursts in BCGs, but our results seem
rather to contradict than to support each other. Therefore this requiress
some discussion.
Their conclusion is drawn from only one point on their
cross-correlation function, corresponding to the minimal sampled distance
of 120~h$^\mathrm{-1}$ kpc. This is about twice the typical
distance  between low-mass companions and BCGs in Taylor et
al.~(\cite{Taylor95}) and the median distance between BCGs and {\it
low-luminosity} neighbours, in our sample (see \ref{subsection:prop}).
The presented uncertainty of this point is so high that
it is difficult to trust any conclusion made on this basis. The
auto-correlation function of the same \ion{H}{ii}-galaxies, determined by
Iovino et al.~(\cite{Iovino88}) and used by Telles \& Maddox~(\cite{Telles00})
as an argument, was measured only for distances larger than
350~h$^\mathrm{-1}$ kpc.
Its extrapolation to the distances of interest, $\sim$50--80~h$^\mathrm{-1}$
kpc, is questionable in the frame of this problem, since its value
is the subject of the study. To indicate possible caveats of their analysis,
we note that about a half of our 32 BCGs with neighbours fainter
than M$_\mathrm{B}$=--18\fm5 (see Table~\ref{Tab_3})
are within the projected distances closer
than 80 kpc (60~h$^{-1}$ kpc). 3 more such faint companions at larger
distances are significantly fainter than M$_\mathrm{B}$=--15, the limit
for APM galaxies at
the typical redshift of their \ion{H}{ii}-galaxies. Therefore, about  60\% of
faint companions found for SBS BCGs would probably not enter the
cross-correlation function of \ion{H}{ii} and APM galaxies, discussed by
Telles \& Maddox.

\subsection{Trigger mechanism and BCG progenitors}
\label{subsection:trigger}

The question of the trigger mechanisms of BCG SF bursts has another
interesting
aspect relating to the evolutionary links of BCGs to other types of low-mass
galaxies. What are BCGs in the periods between the episodes of enhanced SF
activity?
We do not know other gas-rich low-mass galaxies without SF bursts other than
dwarf spirals Sd--Sm, dwarf irregulars, including Im, and LSB dwarfs.
Here we separate LSB dwarfs, as low-mass disks with roughly the same
morphology classification, but with the central blue surface brightness
$\mu_\mathrm{0}$(B) $\geq$ 23\fm5/sq.arcsec (e.g. Dalcanton et
al.~\cite{Dalcanton97}), with the SF rates many
times lower than their `normal' cousins. Unlike more typical
Sdm and dIs, the HI surface density in LSBDs (almost) everywhere is below the
threshold density of gravitational instability, resulting in global
suppression of SF (van der Hulst et al.~\cite{Hulst93}; van Zee et al.
\cite{Zee97}), and their dim optical appearance.

Evidently, either some or all types of gas-rich low-mass galaxies
(in various proportions) can be progenitors of
BCGs/\ion{H}{ii}-galaxies
picked up by ELG surveys thanks to their strong emission lines.
If, as we conclude, the main trigger mechanism of SF bursts in BCGs is
external, then some consequences of BCG in relation to other galaxies
can be derived.

As Mihos et al.~(\cite{Mihos97}) showed in  N-body simulations
of collisions of equal mass high-surface (HSB) and low-surface brightness
(LSB) disk galaxies, their response during a close approach is qualitatively
different. While in the HSB disk a bar structure was generated, which caused
its gas to sink into the center, in the LSB disk only weak spirals were
induced.
Therefore even relatively strong tidal action of a neighbouring galaxy
would cause only a small disturbance of the matter in a LSB galaxy. This
leads to the
conclusion that LSB dwarfs can hardly be progenitors of the main part of
BCGs. Only the merger of LSB dwarfs with any type of low-mass galaxy will
result
in the strong disturbance and the loss of gas stability with subsequent
collapse. In this case, their high HI mass content is the important
factor in providing material for a strong SF burst.

The merging of any type of gas-rich dwarf with an other dwarf will lead to a
strong SF burst and an appearance of the BCG phenomenon. However, among the
field BCGs of merger origin namely the fraction of LSB dwarf progenitors
should prevail. This follows from the fact that the number
density of LSB galaxies is several times higher than that of their HSB
analogs (e.g., Dalcanton et al.~\cite{Dalcanton97}; O'Neil \&
Bothun~\cite{O'Neil00}). This should be the main factor determining the rate
of merger of BCGs, since, unlike the case of weak
tidal trigger, the rate of merging is proportional to the square of the
number density of galaxies under the question. Moreover, being the most
abundant and stable type of galaxy in the general field, LSB dwarfs can
constitute a significant, or even the main fraction of galaxies, exerting
tidal action on BCG progenitors and triggering SF bursts in the majority
of them.

Salzer \& Norton~(\cite{Salzer99}) argued that BCG
progenitors are not typical dwarf irregulars, but rather are objects from
the extreme tail in their distribution on the central mass concentration.
While these authors seemingly appreciate the important feature distinguishing
the essential fraction of BCGs, their BCG sample seems to be randomly
identified, and therefore the claimed results can be biased due to selection
effects. The real fraction of this type of BCG should be confirmed on a BCG
sample with well-defined selection criteria. If, however, Salzer \& Norton are
right and such untypical dIrrs are the progenitors of the majority of
observed BCGs, then such galaxies will be the most easily triggered by
the tidal interactions from external galaxies. These authors suggest an
intrinsic trigger mechanism, based on settling down of the cold ambient gas
ejected from the disk during the previous SF burst. Unlike this,
if interactions are the main trigger mechanism, the SF history,  metal
enrichment, gas consumption and the duty cycle of BCGs should  noticebly
depend on their  environment.
Some evidence for this come from the results of Izotov \& Guseva
(\cite{Izotov89}) of the study of Virgo cluster BCGs. These have higher
metallicities (mean 12+log(O/H) $\sim$8.3) than their typical general field
analogs with a mean of 12+log(O/H)$\sim$8.0 (e.g., Kniazev et
al.~\cite{Kniazev_Gran}). The systematic increase of a BCG parameter
M(HI)/$L_\mathrm{B}$ with the decrease of galaxy density from a cluster
through supercluster and general field to void
environments (Pustilnik et al. \cite{Pustilnik01b}) also supports
the interaction-triggered SF bursts as the main mechanism.

New results of the blind HI survey for low-mass gas-rich galaxies, presented
by Schneider \& Rosenberg~(\cite{Schneider00}), clearly demonstrate a steeply
rising HI mass function of HI-rich {\it field} dwarf galaxies for masses
$<$10$^\mathrm{9}$M$_{\odot}$. This indicates once
more the importance of low-mass companions of field dwarfs (often not seen
without dedicated observations, e.g., Pisano \& Wilcots ~\cite{Pisano99})
as probable triggers of SF activity.
The hypothesis of infall of gas clouds from the intergalactic
medium was recently discussed as an option for triggering  SF bursts in
``isolated'' BCGs (e.g., Crone et al.~\cite{Crone01}). By its essence such
hypothesis is equivalent to the trigger due to sinking satellite if
a gas-rich dwarf galaxy is considered as a massive component. Thus, this
mechanism can in principle trigger SF bursts in some of  BCGs of our sample
without identified disturbing galaxy.

Summarizing the results of this study and the discussion above, we suggest
that BCG progenitors are a mixture of various types of gas-rich low-mass
galaxies with a wide range of surface densities and degrees
of mass concentration.
All of them, being isolated, are in a metastable state with
various levels of ``quasistationary, equilibrium'' star formation
rate.
Some occasionally can ``spontaneousely'' leave it (i.e. due
to some intrinsic processes) and undergo enhanced gas collapse, igniting
SF burst. However, most reach the
phase of SF burst by an external trigger consisting of various strength
interactions with other galaxies. In particular, the
disks with the highest central gas concentration are the least stable against
gas collapse, and can be ingnited to SF burst by relatively weak tidals. On
the other hand, LSB dwarfs are the most stable, and can be transformed to
BCGs only by a merger with another galaxy.

Finally, we note some interesting implications of the
described results for cosmological evolution of gas-rich low-mass galaxies.
If galaxy interactions are the main trigger of enhanced SF
in a significant number of gas-rich galaxies during the modern epoch, then
their role at earlier times becomes dominant due to the significant increase
of galaxy density. This is consistent with much observational data
on the morphology and environment of faint high-$z$ galaxies discovered
with the HST (see, e.g., review by Ferguson et al. \cite{Ferguson00}).
Therefore, for most low-mass gas-rich galaxies with a sufficiently high
surface mass density, such as normal dIs, galaxy interactions during at least
several first Gyr in the life of the Universe seemingly were the main driver
of their cosmological evolution.

The same is applicable to the problem of the existence of local truly young
galaxies. Such objects, if they are not disks of too low surface density,
could survive to the modern epoch as protogalaxies, only if they populated
relatively low galaxy density regions. Therefore, the general field
environment,
or even voids, are the most suitable regions to search for these very
rare objects.

\subsection{Conclusions}

Summarizing the results and discussion presented above we draw the following
conclusions:

\begin{enumerate}
\item{
The fraction of BCGs in the well-defined sample of 86 galaxies which either
have the interacting neighbours with strong enough tidal action or have merger
morphology is $\sim$80\%.
This conclusion remains valid for BCG populating both the Local
Supercluster volume and general field.
}
\item{
Fractions of BCGs with significantly brighter disturbing galaxies
(``non-isolated'' BCGs) vary from  54$\pm$14\% for the Local
Supercluster volume to  $\sim$31.5$\pm$7\% in the general field.
This is consistent with the expectation that large mass
galaxies play a more important role in the regions of higher galaxy
density.
}
\item{
Among  the so called ``isolated'' BCGs (without bright
neighbouring galaxy) in both the Local Supercluster volume and in the
general field, $\sim$43$\pm$10\% are disturbed by dwarf galaxies and
$\sim$26$\pm$8\% have a merger morphology.
}
\item{
The whole BCG dataset discussed in the paper implies that galaxy interactions
with both massive and dwarfs neighbours and with the full range of the
disturbance
amplitude (from weak tidals to mergers), is very important, and probably the
main mechanism that triggers SF bursts in BCG progenitors. These
are the mixture of all known low-mass gas-rich galaxy types with proportions
which still are unknown. Depending on the quiescent level of SF
of various BCG progenitors, their stability against various
interaction-induced disturbances and their type of environment, interactions
may be the main driver of their cosmological evolution.
Future high quality N-body simulations of gas collapse in gas-rich galaxies
due to the tidal action of external galaxies are necessary in order
to support more certain conclusions.
}
\end{enumerate}


\begin{acknowledgements}

We are pleased to thank T.Kniazeva for the help in the reduction of
observational data and D.Makarov for consultations. Our thanks to
K.Noeske for sending us his and co-authors' article prior publication,
and to J.Schombert, who kindly provided us with some unpublished data
on the LSB galaxy sample.
The authors thank the anonymous referee for useful suggestions.
We acknowledge the partial support from Russian state program
"Astronomy" and Center of Cosmoparticle Physics "Cosmion". This research has
made use of the NASA/IPAC Extragalactic
Database (NED) which is operated by the Jet Propulsion Laboratory,
California Institute of Technology, under contract with the National
Aeronautics and Space Administration. The use of the Digitized Sky Survey
(DSS-II) and APM Database is gratefully acknowledged.   \\

\end{acknowledgements}

\renewcommand{\baselinestretch}{0.5}

\clearpage


\begin{table*}[h]
\centering{
\caption{Main parameters of the studied BCG sample}
\label{Tab_2}
{
\begin{tabular}{rlllrrrrrll} \hline
\#&
\MC{1}{c}{BCG Name}            &
\MC{1}{c}{ $\alpha_\mathrm{1950.0}$ } &
\MC{1}{c}{ $\delta_\mathrm{1950.0}$ } &
\MC{1}{c}{ V$_\mathrm{hel}$}   &
\MC{1}{c}{ Ref}                &
\MC{1}{c}{ D$_\mathrm{Vir}$ }  &
\MC{1}{c}{ $m_\mathrm{B}$ }    &
\MC{1}{c}{ M$_\mathrm{B}$}     &
\MC{1}{c}{ Glob.}              &
\MC{1}{c}{ Other}              \\
&         &
\MC{1}{c}{ $h$~$m$~$s$ }                     &
\MC{1}{c}{ $\circ$~$\prime$~$\prime\prime$ } &
\MC{1}{c}{km/s}                              &
\MC{1}{c}{Vel }                              &
\MC{1}{c}{ Mpc }                             &
\MC{1}{c}{ $mag$ }                           &
\MC{1}{c}{ $mag$ }                           &
\MC{1}{c}{ envir }                           &
\MC{1}{c}{ Names }                           \\
&
\MC{1}{c}{ (1) } &
\MC{1}{c}{ (2) } &
\MC{1}{c}{ (3) } &
\MC{1}{c}{ (4) } &
\MC{1}{c}{ (5) } &
\MC{1}{c}{ (6) } &
\MC{1}{c}{ (7) } &
\MC{1}{c}{ (8) } &
\MC{1}{c}{ (9) } &
\MC{1}{c}{ (10) } \\
\hline
\qq&0741+535 & 07 41 34.8 &+53 32 55 & 5564 & 1 & 77.1 &*17.06& --17.38 &GF& Mkn~1409 \\
\qq&0749+568 & 07 49 37.8 &+56 49 41 & 5488 & 1 & 76.3 &+18.02& --16.39 &GF& \\
\qq&0834+518 & 08 34 03.6 &+51 48 53 &  749 & 2 & 13.0 &*17.22& --13.35 &LS& Mkn~94 \\
\qq&0912+599 & 09 12 53.5 &+59 58 53 & 4150 & 1 & 59.2 &+15.39& --18.47 &GF& Mkn~19 \\
\qq&0913+536 & 09 13 12.7 &+53 39 07 & 2203 & 2 & 33.8 &n14.60& --18.04 &GF& Mkn~104 \\
\qq&0917+527 & 09 17 25.7 &+52 46 53 & 2323 & 1 & 35.4 & 16.48& --16.27 &GF& Mkn~1416 \\
\qq&0926+606A& 09 26 20.1 &+60 40 02 & 4002 & 1 & 59.1 &+16.77& --17.09 &GF& \\
\qq&0926+606B& 09 26 23.0 &+60 41 15 & 4090 & 1 & 59.1 &+17.31& --16.55 &GF& \\
\qq&0930+554 & 09 30 30.3 &+55 27 46 &  751 & 1 & 14.2 &+16.50& --14.26 &LS& Mkn~116=IZw18 \\
\qq&0940+544 & 09 40 51.0 &+54 25 21 & 1638 & 1 & 26.7 &+17.32& --14.81 &GF& \\
\qq&0942+573 & 09 42 56.4 &+57 20 55 & 1177 & 1 & 20.6 &*17.17& --14.40 &LS& Mkn~1424 \\
\qq&0943+543 & 09 43 40.7 &+54 19 33 & 1644 & 2 & 26.8 &+17.37& --14.79 &GF& \\
\qq&0946+558 & 09 46 03.4 &+55 48 49 & 1551 & 1 & 25.6 & 16.28& --15.76 &LS& Mkn~22 \\
\qq&1001+555 & 10 01 20.7 &+55 33 22 & 1116 & 1 & 19.7 & 16.02& --15.45 &LS& \\
\qq&1008+589 & 10 08 31.7 &+58 58 53 & 2096 & 2 & 32.9 &*16.19& --16.40 &GF& Mkn~27 \\
\qq&1011+601 & 10 11 31.5 &+60 06 07 & 2160 & 1 & 33.7 &+17.23& --15.41 &GF& \\
\qq&1011+600 & 10 11 33.7 &+60 03 59 & 2160 & 1 & 33.7 & 17.26& --15.38 &GF& \\
\qq&1029+546 & 10 29 22.7 &+54 39 30 & 1451 & 2 & 24.3 &n13.40& --18.55 &LS& Mkn~33 \\
\qq&1030+583 & 10 30 56.3 &+58 19 20 & 2269 & 2 & 35.1 & 16.82& --16.03 &GF& Mkn~1434 \\
\qq&1033+531 & 10 33 30.6 &+53 06 34 &  986 & 1 & 17.9 & 16.16& --15.10 &LS& \\
\qq&1037+494 & 10 37 42.1 &+49 28 05 & 1496 & 1 & 25.0 & 16.49& --15.50 &LS& \\
\qq&1042+562 & 10 42 16.0 &+56 13 21 &  944 & 2 & 17.4 & 13.34& --17.86 &LS& Mkn~35 \\
\qq&1051+545 & 10 51 38.2 &+54 34 18 & 1350 & 2 & 23.2 & 12.45& --19.39 &LS& Arp~205 \\
\qq&1054+504 & 10 54 02.7 &+50 24 28 & 1354 & 1 & 23.4 & 16.20& --15.65 &LS& \\
\qq&1057+511B& 10 57 10.6 &+51 07 30 & 3056 & 2 & 45.2 & 15.99& --17.32 &GF& \\
\qq&1114+587 & 11 14 10.7 &+58 47 22 & 1594 & 1 & 26.7 &+16.33& --15.80 &GF& \\
\qq&1114+517 & 11 14 42.7 &+51 42 16 & 2838 & 1 & 43.4 & 16.41& --16.78 &GF& Mkn~1445 \\
\qq&1116+517 & 11 16 45.2 &+51 46 38 & 1341 & 2 & 23.1 & 17.14& --14.68 &LS& CG~1116+51 \\
\qq&1119+586 & 11 19 45.7 &+58 36 05 & 1583 & 1 & 26.5 &+18.56& --13.56 &LS& \\
\qq&1122+575 & 11 22 12.8 &+57 32 37 & 1336 & 3 & 23.2 &*17.51& --14.32 &LS& \\
\qq&1123+576 & 11 23 23.2 &+57 37 42 &  289 & 1 &3.5$*$& 16.76& --10.96 &LS& \\
\qq&1123+594 & 11 23 53.7 &+59 25 50 & 1199 & 2 & 21.4 & 14.44& --17.21 &LS& Mkn~169 \\
\qq&1124+541 & 11 24 45.5 &+54 11 26 & 2940 & 1 & 44.0 & 16.17& --17.05 &GF& Mkn~1446 \\
\qq&1125+562 & 11 25 16.8 &+56 13 03 & 5591 & 2 & 78.0 & 16.82& --17.64 &GF& \\
\qq&1128+573 & 11 28 29.1 &+57 20 33 & 1799 & 2 & 29.1 & 18.18& --14.12 &GF& \\
\qq&1129+576 & 11 29 15.5 &+57 39 17 & 1566 & 2 & 26.3 & 16.50& --15.60 &LS& \\
\qq&1130+495 & 11 30 45.7 &+49 30 54 &  249 & 2 & 3.5  &n14.50& --13.22 &LS& Mkn~178 \\
\qq&1135+581 & 11 35 51.3 &+58 09 24 &  946 & 2 & 18.1 & 15.85& --15.44 &LS& Mkn~1450 \\
\qq&1136+607 & 11 36 26.3 &+60 47 22 & 3457 & 2 & 51.0 &*17.74& --15.80 &GF& \\
\qq&1137+589 & 11 37 48.4 &+58 55 10 & 2032 & 1 & 32.5 & 18.35& --14.21 &GF& \\
\qq&1147+520 & 11 47 16.4 &+52 00 51 & 1079 & 2 & 19.4 & 17.03& --14.70 &LS& \\
\qq&1149+596 & 11 49 55.9 &+59 39 37 & 3351 & 1 & 49.0 &*15.98& --17.47 &GF& \\
\qq&1152+579 & 11 52 51.9 &+57 56 34 & 5157 & 1 & 72.6 & 17.08& --17.22 &GF& Mkn~193 \\
\qq&1159+545 & 11 59 28.9 &+54 32 32 & 3537 & 2 & 51.6 & 18.26& --15.36 &GF& \\
\qq&1159+516B& 11 59 31.7 &+51 40 18 & 4424 & 1 & 63.2 & 17.23& --16.77 &GF& \\
%
%
%
%
\qq&1203+592 & 12 03 43.2 &+59 14 59 & 3256 & 1 & 48.1&*16.23 & --17.18 &GF& \\
\qq&1205+557 & 12 05 57.6 &+55 42 08 & 1740 & 1 & 28.7& 17.35 & --14.94 &GF& \\
\qq&1211+540 & 12 11 34.0 &+54 01 59 &  907 & 1 & 17.2& 17.67 & --13.51 &LS& \\
\qq&1212+493 & 12 12 06.1 &+49 23 28 & 3607 & 2 & 55.0&*17.64 & --16.06 &GF& \\
\qq&1213+597 & 12 13 34.9 &+59 47 09 & 4420 & 1 & 62.7& 15.48 & --18.51 &GF& Mkn~1468 \\
\qq&1221+585 & 12 21 19.7 &+58 31 23 & 4350 & 3 & 62.1& 16.52 & --17.45 &GF& \\
\qq&1221+545B& 12 21 58.6 &+54 31 24 & 5704 & 1 & 78.1& 18.26 & --16.20 &GF& \\
\qq&1222+588 & 12 22 29.2 &+58 49 17 & 4722 & 2 & 67.1& 17.07 & --17.06 &GF& \\
\qq&1222+614 & 12 22 44.3 &+61 25 42 &  734 & 2 & 14.1& 15.21 & --15.54 &LS& \\
\qq&1241+549 & 12 41 30.3 &+54 55 11 & 4945 & 1 & 67.5&*17.92 & --16.23 &GF& \\
\qq&1242+549 & 12 42 20.2 &+54 59 51 & 4740 & 2 & 67.0& 15.93 & --18.21 &GF& \\
\qq&1307+563 & 13 07 41.7 &+56 18 23 & 5165 & 2 & 73.1&+18.29 & --16.03 &GF& \\
\qq&1314+605 & 13 14 54.1 &+60 34 21 & 1848 & 1 & 30.3& 17.10 & --15.31 &GF& \\
\qq&1317+523A& 13 17 42.8 &+52 19 57 & 4619 & 1 & 65.8&*16.22 & --17.87 &GF& \\
\qq&1318+520 & 13 18 07.6 &+52 01 37 & 4740 & 1 & 67.4& 15.92 & --18.22 &GF& \\
\qq&1319+579 & 13 19 27.0 &+57 57 22 & 1978 & 2 & 33.5& 15.55 & --17.08 &GF& NGC~5113  \\
\qq&1331+493 & 13 31 19.9 &+49 21 32 &  590 & 1 & 11.3& 15.11 & --15.16 &LS& \\
\qq&1340+529 & 13 40 56.0 &+52 57 22 & 1801 & 1 & 29.6& 16.36 & --16.00 &GF& Mkn~1480 \\
\qq&1341+529 & 13 41 03.7 &+52 56 22 & 1797 & 1 & 29.7& 16.59 & --15.77 &GF& Mkn~1481 \\
\qq&1341+594 & 13 41 45.1 &+59 25 07 & 3037 & 1 & 45.6& 17.48 & --15.81 &GF& \\
\qq&1358+554E& 13 58 48.8 &+55 29 44 & 3857 & 1 & 55.9&*16.46 & --17.28 &GF& \\
\qq&1401+490 & 14 01 50.0 &+49 00 18 &  777 & 1 & 14.9& 17.28 & --13.59 &LS& CG~337 \\
\qq&1413+495 & 14 13 18.7 &+49 34 26 & 3923 & 1 & 56.8& 16.82 & --16.95 &GF& CG~383 \\
\qq&1423+517 & 14 23 40.9 &+51 46 45 & 2141 & 2 & 34.1& 16.80 & --15.86 &GF& CG~424 \\
\qq&1430+596 & 14 30 03.5 &+59 36 42 & 1855 & 1 & 30.4& 16.44 & --15.97 &GF& CG~470 \\
\qq&1430+526A& 14 30 53.3 &+52 37 59 & 3424 & 1 & 50.4&*16.30 & --17.21 &GF& \\
\qq&1446+595 & 14 46 20.6 &+59 34 20 & 2121 & 2 & 33.9& 17.15 & --15.50 &GF& \\
\qq&1448+526 & 14 48 27.6 &+52 36 43 &  726 & 1 & 14.9& 15.76 & --15.11 &LS& Mkn~826 \\
\qq&1453+526 & 14 53 09.2 &+52 37 52 & 3267 & 2 & 48.1& 17.08 & --16.33 &GF& CG~577 \\
\qq&1503+531 & 15 03 02.0 &+53 06 54 & 5619 & 2 & 78.6& 17.26 & --17.22 &GF& CG~618 \\
\qq&1504+514N& 15 04 18.8 &+51 26 54 & 3796 & 1 & 55.1& 15.11 & --18.60 &GF& CG~625 \\
\qq&1523+519 & 15 23 51.6 &+51 55 22 & 3645 & 1 & 53.1& 17.05 & --16.58 &GF& CG~707 \\
\qq&1524+554 & 15 24 40.1 &+55 25 30 & 3409 & 1 & 50.3&*16.83 & --16.68 &GF& \\
\qq&1533+574 & 15 33 03.8 &+57 27 01 & 3310 & 1 & 49.0& 16.23 & --17.22 &GF& \\
\qq&1535+554 & 15 35 48.8 &+55 25 35 &  667 & 1 & 13.9& 15.44 & --15.28 &LS& Mkn~487 \\
\qq&1551+601A& 15 51 46.4 &+60 11 29 & 2932 & 1 & 44.2& 16.68 & --16.55 &GF& \\
\qq&1555+515 & 15 55 40.2 &+51 30 54 & 3617 & 3 & 55.3& 18.51 & --15.20 &GF& \\
\qq&1620+577 & 16 20 15.4 &+57 44 45 & 4848 & 4 & 60.6& 16.75 & --17.16 &GF& \\
\qq&1632+578 & 16 32 40.8 &+57 52 38 & 5530 & 1 & 69.7& 18.09 & --16.13 &GF& \\
\qq&1634+523 & 16 34 07.8 &+52 18 57 & 2700 & 1 & 41.0& 15.84 & --17.22 &GF& Mkn~1499 \\
\qq&1707+565 & 17 07 03.4 &+56 34 34 & 3323 & 1 & 49.3& 16.39 & --17.07 &GF& \\
\hline \\[-0.2cm]
\end{tabular}
}
}
\end{table*}

\clearpage

\begin{table*}[h]
\setcounter{qub}{0}
\centering{
\caption{Parameters of the strongest disturbers of BCGs -- probable SF burst triggers}
\label{Tab_3}
{
\begin{tabular}{rllllrrrrlrrl} \hline
\#&
\MC{1}{c}{BCG name}     &
\MC{1}{c}{Partner's}    &
\MC{1}{c}{ $\alpha_\mathrm{1950.0}^\mathrm{Part}$ } &
\MC{1}{c}{ $\delta_\mathrm{1950.0}^\mathrm{Part}$ } &
\MC{1}{c}{V$_\mathrm{hel}^\mathrm{BCG}$} &
\MC{1}{c}{V$_\mathrm{hel}^\mathrm{Part}$}&
\MC{1}{c}{Ref   }     &
\MC{1}{c}{Part.   }     &
\MC{1}{c}{Part.   }     &
\MC{1}{c}{$\Delta$M$_\mathrm{B}$}&
\MC{1}{c}{D$_\mathrm{Proj}$   }     &
\MC{1}{c}{Trig    }     \\
&                       &
\MC{1}{c}{name}         &
\MC{1}{c}{ $h$~$m$~$s$ }       &
\MC{1}{c}{ $\circ$~$\prime$~$\prime\prime$ } &
\MC{1}{c}{km/s         }&
\MC{1}{c}{km/s         }&
\MC{1}{c}{V            }&
\MC{1}{c}{B$_\mathrm{NED}$    }&
\MC{1}{c}{M$_\mathrm{B}^\mathrm{T}$    }&
			&
\MC{1}{c}{ kpc         }&
\MC{1}{c}{type         }\\
		 &
\MC{1}{c}{ (1) } &
\MC{1}{c}{ (2) } &
\MC{1}{c}{ (3) } &
\MC{1}{c}{ (4) } &
\MC{1}{c}{ (5) } &
\MC{1}{c}{ (6) } &
\MC{1}{c}{ (7) } &
\MC{1}{c}{ (8) } &
\MC{1}{c}{ (9) } &
\MC{1}{c}{ (10)} &
\MC{1}{c}{ (11)}  \\
 \hline
\qq&0741+535  &0741+535   &  --        &  --      &5462  & -- &  & --    & --   & --          &  --  & m?  \\
\qq&0749+568  & --        &  --        &  --      &5471  & -- &  & --    & --   & --          &  --  &     \\
\qq&0834+518  &UGC 4499   &08 34 02.0  &51 49 38  & 749  & 691& 5& 13.5  &--16.3& --2.9       &   3  & p   \\
\qq&0912+599  &0912+599   &  --        &  --      &4150  & -- &  & --    & --   & --          &  --  & m   \\
\qq&0913+536  &UGC 4906   &09 14 08.1  &53 12 16  &2203  &2307& 5& 13.5  &--19.0&--1.0        & 277  & b   \\
\qq&0917+527  &UGC 4906   &09 14 08.1  &53 12 16  &2323  &2307& 5& 13.5  &--19.0&--2.7        & 404  & p  \\
\qq&0926+606A &S0926+606B &09 26 23.0  &60 41 15  &4002  &4090& 1&+17.3  &--16.4&+0.7         &  22  & b   \\
\qq&0926+606B &S0926+606A &09 26 20.1  &60 40 02  &4090  &4002& 1&+16.8  &--16.8&--0.2        &  22  & b   \\
\qq&0930+554  & --        &  --        &  --      & 751  & -- &  & --    & --   & --          &  --  &    \\
\qq&0940+544  &S0943+543  &09 43 40.7  &54 19 33  &1638  &1644& 2&+17.4  &--14.3&+0.1         & 197  & b   \\
\qq&0942+573  &0942+573   & --         &  --      &1177  & -- &  & --    & --   & --          & --   & m  \\
\qq&0943+543  &S0940+544  &09 40 51.0  &54 25 21  &1644  &1638&  &+17.3  &--14.4&--0.1        & 198  & b   \\
\qq&0946+558  &0946+558   & --         &  --      &1551  & -- &  & --    & --   & --          & --   & m?  \\
\qq&1001+555  &NGC 3079   &09 58 35.0  &55 55 15  &1116  &1125& 2& 11.5  &--19.4&--3.9        & 184  & p   \\
\qq&1008+589  &UGC 5480   &10 06 48.9  &58 44 07  &2096  &2145& 2& 15.4  &--16.9&--0.5        & 190  & b   \\
\qq&1011+600  &S1011+601  &10 11 31.5  &60 06 07  &2160  &2160& 1&+17.2  &--15.1&+0.3         &  21  & b   \\
\qq&1011+601  &S1011+600  &10 11 33.7  &60 03 59  &2160  &2160& 1&+17.3  &--15.0&+0.4         &  21  & b   \\
\qq&1029+546  &UGC 5676   &10 25 53.7  &54 58 19  &1451  &1412& 5& 14.5  &--16.9&+1.6         & 251  & f   \\
\qq&1030+583  &1030+583   & --         & --       &2269  & -- &  & --    & --   & --          &  --  & m?  \\
\qq&1033+531  &NGC 3310   &10 35 40.1  &53 45 49  & 986  & 983& 5& 11.0  &--19.6&--4.5        & 228  & p   \\
\qq&1037+494  & --        &  --        & --       &1496  & -- &  & --    & --   & --          &  --  &     \\
\qq&1042+562  &UGC 5888   &10 44 40.3  &56 21 17  & 944  &1239& 2& 14.8  &--16.3&+1.6         & 109  & f   \\
\qq&1051+545  &UGC 6016   &10 51 12.7  &54 33 12  &1350  &1493& 2&*15.9  &--15.6&+3.8         &  26  & f  \\
\qq&1054+504  &UGC 6029   &10 52 06.1  &49 59 36  &1375  &1363& 2& 14.0  &--17.3&--1.6        & 211  & p  \\
\qq&1057+511B &UGC 6074   &10 57 02.6  &51 10 08  &3056  &2885& 1& 14.2  &--18.7&--1.4        &  38  & p  \\
\qq&1114+587  &UGC 6304   &11 14 55.7  &58 37 33  &1594  &1762& 2&*15.9  &--16.0&--0.2        &  89  & b  \\
\qq&1114+517  &UGC 6309   &11 14 56.7  &51 45 01  &2838  &2870& 1& 13.7  &--19.2&--2.4        &  44  & p  \\
\qq&1116+517  & --        & --         & --       &1341  & -- &  & --    & --   & --          & --   &    \\
\qq&1119+586  &NGC 3642   &11 19 25.6  &59 21 01  &1583  &1588& 2& 11.9  &--19.8&--6.2        & 346  & p  \\
\qq&1122+575  &NGC 3683   &11 24 43.1  &57 09 14  &1336  &1716& 2& 13.2  &--18.6&--4.3        & 209  & p  \\
\qq&1123+576  &NGC 3738   &11 33 04.5  &54 47 58  & 289  & 229& 2& 11.8  &--15.6&--4.6        & 191  & p  \\
\qq&1123+594  &NGC 3642   &11 19 25.0  &59 20 55  &1199  &1588& 2& 11.9  &--19.8&--2.6        & 215  & p  \\
\qq&1124+541  &NGC 3656   &11 20 50.5  &54 07 08  &2940  &2869& 2& 13.3  &--19.6&--2.6        & 445  & p  \\
\qq&1125+562  &S1124+561  &11 24 54.1  &56 11 47  &5591  &5456& 6&*15.9  &--18.4&--0.8        &  77  & b  \\
\qq&1128+573  &NGC 3683   &11 24 43.1  &57 09 14  &1799  &1716& 2& 13.2  &--18.6&--4.5        & 275  & p  \\
\qq&1129+576  &S1129+577  &11 29 16.7  &57 43 00  &1566  &1457& 2&*15.9  &--15.5&+0.1         &  29  & b  \\
\qq&1130+495  &1130+495   & --         &  --      & 249  & -- &  & --    & --   & --          & --   & m  \\
\qq&1135+581  &UGC 6616   &11 36 36.9  &58 32 43  & 946  &1154& 2& 13.7  &--17.2&--1.8        & 127  & p  \\
\qq&1136+607  &UGC 6619   &11 36 43.4  &60 49 26  &3457  &3465& 5& 15.4  &--17.9&--2.1        &  44  & p  \\
\qq&1137+589  &1137+589   & --         & --       &2032  & -- &  &  --   & --   & --          &  --  & m? \\
\qq&1147+520  &NGC 3917   &11 48 07.7  &52 06 09  &1079  & 968& 2& 12.5  &--18.1&--3.4        &  54  & p  \\
\qq&1149+596  &NGC 3894   &11 46 11.4  &59 41 41  &3351  &3223& 2& 12.6  &--20.6&--3.1        & 405  & p  \\
\qq&1152+579  &UGC 6939   &11 55 07.4  &57 50 38  &5157  &4987& 5& 14.6  &--19.5&--2.3        & 400  & p  \\
\qq&1159+545  &1159+545   & --         & --       &3537  & -- &  & --   & --    & --          &  --  & m? \\
\qq&1159+516B & --        & --         & --       &4424  & -- &  & --   &  --   & --          &  --  &    \\
%
%
\qq&1203+592  &1203+592   & --         & --       &3256  & -- &  & --   & --   &--          &  --  & m  \\
\qq&1205+557  & --        & --         & --       &1740  & -- &  & --   & --   &--          &  --  &    \\
\qq&1211+540  &NGC 4142   &12 07 00.6  &53 22 58  & 907  &1157& 2&13.9  &--16.3& --2.8      & 282  & p  \\
\qq&1212+493  & --        & --         & --       &3607  & -- &  & --   & --   & --         &  --  &    \\
\qq&1213+597  &NGC 4195   &12 11 51.0  &59 53 32  &4420  &4350& 5& 14.9 &--18.9&--0.4       & 265  & b  \\
\qq&1221+585  &NGC 4358   &12 21 39.1  &58 39 44  &4350  &4497& 2& 14.3 &--19.6&--2.1       & 158  & p  \\
\qq&1221+545B & --        & --         & --       &5704  & -- &  & --   & --   &--          &  --  &    \\
\qq&1222+588  &NGC 4358   &12 21 39.1  &58 39 44  &4722  &4497& 2& 14.3 &--19.6&--2.5       & 225  & p  \\
\qq&1222+614  &M10-18-044 &12 22 32.6  &61 20 26  & 734  & 709& 2&*15.3 &--14.6&+0.9        &  23  & b  \\
\qq&1241+549  &Mkn 220    &12 41 31.8  &55 10 11  &4945  &4933& 2& 14.1 &--20.0&--3.8       & 295  & p  \\
\qq&1242+549  &A1242+5458 &12 42 24.4  &54 58 48  &4740  &4598& 4&+18.6 &--15.3& +2.9       &  26  & f  \\
\qq&1307+563  & --        &  --        &  --      &5165  & -- &  & --   & --   & --         &  --  &    \\
\qq&1314+605  &UGC 8282   &13 09 53.9  &60 30 46  &1848  &2073& 5& 14.8 &--17.4&--2.1       & 327  & p  \\
\qq&1317+523A &A1317+5219 &13 17 44.1  &52 19 24  &4619  &4737& 4&*16.6 &--17.4&+0.5        &  12  & b  \\
\qq&1318+520  & --        &  --        &  --      &4740  & -- &  & --   & --   & --         & --   &    \\
\qq&1319+579  &NGC 5109   &13 18 55.3  &57 54 16  &1978  &2131& 5& 13.8 &--18.8&--1.7       &  52  & p  \\
\qq&1331+493  &NGC 5194   &13 27 46.0  &47 27 22  & 590  & 463& 2&  9.0 &--20.0&--4.8       & 393  & p  \\
\qq&1340+529  &Mkn 1481   &13 41 03.5  &52 56 22  &1801  &1797& 1&+16.6 &--15.3& +0.7       &  13  & b  \\
\qq&1341+529  &Mkn 1480   &13 40 55.9  &52 57 36  &1797  &1801& 1&+16.4 &--15.5& +0.3       &  13  & b  \\
\qq&1341+594  & --        & --         &  --      &3037  & -- &  & --   & --   & --         &  --  &    \\
\qq&1358+554E &1358+554E  & --         &  --      &3857  & -- &  & --   & --   & --         &  --  & m  \\
\qq&1401+490  & --        & --         &  --      & 777  & -- &  & --   & --   & --         &  --  &    \\
\qq&1413+495  & --        & --         &  --      &3923  & -- &  & --   & --   & --         &  --  &    \\
\qq&1423+517  &NGC 5624   &14 24 51.8  &51 48 30  &2142  &1923& 2& 14.1 &--17.9&--2.0       & 110  & p  \\
\qq&1430+596  &NGC 5667   &14 28 28.2  &59 41 29  &1855  &1943& 2& 13.2 &--18.9&--2.9       & 115  & p  \\
\qq&1430+526A & --        & --         & --       &3424  & -- &  & --   & --   & --         &  --  &    \\
\qq&1446+595  &NGC 5777   &14 49 59.8  &59 10 58  &2121  &2145& 2& 14.1 &--18.2&--2.7       & 358  & p  \\
\qq&1448+526  &UGC 9630   &14 45 51.7  &53 02 21  & 729  & 716& 2& 15.9 &--14.0&+1.1        & 151  & b  \\
\qq&1453+526  & --        & --         &  --      &3267  & -- &  &  --  & --   & --         &  --  &    \\
\qq&1503+531  &UGC 9688   &15 02 32.9  &53 07 22  &5619  &5833& 5& 14.7 &--19.8&--2.6       & 100  & p  \\
\qq&1504+514N &UGC 9702   &15 03 41.0  &51 21 03  &3796  &3855& 4& 15.6 &--18.0&+0.6        & 133  & b  \\
\qq&1523+519  &  --       &  --        &  --      &3645  & -- &  & --   & --   & --         &  --  &    \\
\qq&1524+554  &Mkn 481    &15 26 33.2  &55 36 33  &3409  &3298& 2&*16.7 &--16.5&+0.2        & 285  & b  \\
\qq&1533+574  &1533+574   & --         & --       &3310  & -- &  & --   & --   & --         &  --  & m  \\
\qq&1535+554  &NGC 5963   &15 32 16.0  &56 43 31  & 665  & 655& 2& 13.0 &--16.7&--1.4       & 337  & p  \\
\qq&1551+601A &1551+601A  & --         & --       &2932  & -- &  & --   & --   & --         & --   & m? \\
\qq&1555+515  &UGC 10123  &15 57 41.1  &51 26 43  &3617  &3706& 5& 15.1 &--18.4&--3.2       & 309  & p  \\
\qq&1620+577  &NGC 6130   &16 18 34.6  &57 43 56  &4848  &5137& 5& 14.3 &--19.9&--2.7       & 238  & p  \\
\qq&1632+578  &NGC 6187   &16 30 41.0  &57 48 43  &5530  &5477& 5&*15.3 &--19.0&--2.9       & 333  & p  \\
\qq&1634+523  &1634+523   & --         &  --      &2700  & -- &  & --   & --   & --         & --   & m  \\
\qq&1707+565  &1707+565   & --         &  --      &3323  & -- &  & --   & --   & --         &  --  & m? \\
\hline \\[-0.2cm]
\end{tabular}
}
}
\end{table*}

\clearpage
{\bf Notes for Table~\ref{Tab_3}}             \\
{\bf 0741+535} well advanced merger morphology. Bright galaxy NGC 2431 with
 M$_B=-$22.4 at the projected distance 480 kpc and $\Delta$V=216 km~s$^{-1}$  \\
{\bf 0749+568} faint galaxy ($\Delta$B$\sim$2$^m$) at $\sim$23$''$ to SW.  \\
{\bf 0834+518} very close in projection to much more massive galaxy.  \\
{\bf 0912+599} two disks in contact, both with SF burst  \\
{\bf 0926+606A/0926+606B} interacting pair of BCGs with synchronized SF burst \\
{\bf 0930+554 } complex morphology with propagating SF wave (Izotov 1999, private communication) \\
{\bf 0940+544/0943+543} interacting pair of BCGs. \\
{\bf 0942+573 } merger/galaxy pair (Mazzarella et al.~\cite{Mazza91})   \\
{\bf 0946+558 } double nuclei galaxy (Mazzarella et al.~\cite{Mazza91}) \\
{\bf 1001+555 } bent edge-on disk. \\
{\bf 1011+601A/1011+600B} interacting pair of BCGs with synchronized SF burst. \\
{\bf 1029+546 } There is a galaxy of $\sim$3$^m$ fainter
  at 4.7$^{\prime}$ to NEE. If at the same distance, its
  projected distance from Mkn~33 is $\sim$30 kpc. \\
{\bf 1030+583 } Disturbed morphology. \\
{\bf 1051+545 } Faint companion with evidences of interaction (Karachentseva
et al.~\cite{Kara96}). \\
{\bf 1057+511B} Highly disturbed morphology. \\
{\bf 1116+517 } There are galaxies with M$_B=-18.7$ and $-20.8$ at $\sim$600 kpc. \\
{\bf 1119+586 } Besides of massive companion NGC~3642, there is another
 massive galaxy NGC~3619 with close parameters and projected distance,
 and two more probable companions: SBS~1121+586 at 74 kpc and $\sim$1\fm6
 brighter and SBS~1119+583 at 139 kpc and $\sim$1\fm8 brighter. They all
 probably  enter to one group. \\
{\bf 1123+576 } Member of close complex in Canes Venatici. For
   brighter dwarf companion NGC~3738 photometrical distance 3.52 Mpc
   from Georgiev et al.~(\cite{Georgiev97}) is accepted. \\
{\bf 1128+573 } Massive parent. Merger morphology.      \\
{\bf 1129+576 } Brighter dwarf companion with bent
   edge-on disk.    \\
{\bf 1130+495 } Member of close complex in Canes Venatici.
   Merger, or galaxy pair ? (Mazzarella et al.~\cite{Mazza91})    \\
{\bf 1135+581 } Brighter dwarf companion with SF burst in
   very center.    \\
{\bf 1149+596 } Companion of galaxy pair KPG 303 ?   \\
{\bf 1159+545 } Merger, or companion of massive spiral Mkn~433  \\
{\bf 1159+516B} Pair of close dwarfs ? No velocity for candidate companion \\
{\bf 1203+592 } Very close faint companion.                 \\
{\bf 1205+557 } SF burst in center and disturbed morphology. Several galaxies
  of comparable brightness in the vicinity with no redshift.    \\
{\bf 1211+540 } Brighter dwarf companion. BCG has disturbed  morphology.  \\
{\bf 1212+493 } No companions ? Massive galaxy with very close V$_r$ at
   $\sim$700 kpc.                      \\
{\bf 1213+597 } Disturbed morphology.         \\
{\bf 1221+545B} No candidate companions.                        \\
{\bf 1222+588 } NGC 4335 at 280 kpc is one more massive companion with
  comparable tidal effect.  BCG has disturbed perephery. \\
{\bf 1242+549 } BCG with tidal tail. Faint blue companion at $\sim$25 kpc. \\
{\bf 1307+563 } 2 candidate brighter companions with no velocity.  \\
{\bf 1314+605 } Brighter dwarf companion.     \\
{\bf 1317+523A} There is a third galaxy Mkn~251 of close brightness at
  $\sim$60 kpc and $\Delta$V=-60 km~s$^{-1}$.                \\
{\bf 1318+520 } Close edge-on disk galaxy at 2' with no velocity. \\
{\bf 1331+493 } Massive parent. BCG with propogating SF (Zenina et al.~\cite{Zenina97}).  \\
{\bf 1340+529/1341+529 } Close pair of BCGs with synchronized SF burst.  \\
{\bf 1341+594 } No evident companions. Disturbed  morphology. \\
{\bf 1358+554E} Disturbed morphology, probable merger.               \\
{\bf 1413+495 } No neighbours ? Highly disturbed morphology. The optical
 redshift, consistent with HI one from Thuan et al.~(\cite{Thuan99}) is
 presented at first time. That one in Doublier et al.~(\cite{Doublier97})
 is mistaken. \\
{\bf 1430+526A} No neighbours ? Irr. morphology. Two close candidate
  companions  with no velocity.     \\
{\bf 1503+531 } Massive parent. BCG has disturbed morphology.      \\
{\bf 1523+519 } No neighbours ? Faint galaxy at $\sim$3' with no velocity. \\
{\bf 1524+554 } One more brighter galaxy Mkn 482,  2.1
mag. brighter, at 370 kpc with similar tidal effect. \\
{\bf 1533+574 } Merger, or a massive companion at $\sim$500 kpc.      \\
{\bf 1551+601A} Merger, or pair with a similar galaxy at $\sim$300 kpc.  \\
{\bf 1634+523 } = I Zw 159. Blue double compact with faint blue plumes north and south. (NED).  \\

\renewcommand{\baselinestretch}{1.0}

\clearpage

\appendix
\section{Spectral observations and results}

\subsection{Observations and reduction}

All observations have been conducted with the SAO RAS 6\,m telescope
during the period 1996--2000. Besides candidate companion galaxies,
three SBS BCGs were also observed. For one of them (SBS~1413+495) the
redshift was unknown before, and for two others we tried to improve the
accuracy of redshifts to make more confident our search for companions.
This optical redshift for SBS~1413+495 was later confirmed and its
precision was improved using HI data by Thuan et al.~(\cite{Thuan99}).
Candidate neighbours for two
additional BCGs not entering into the sample of 86 BCGs were also observed,
and appeared to be real companions (SBS 0916+542 and 1120+586). Observational
results on these two new galaxies are given as well in Table A1.

Three set-ups were used for the observations:
\begin{enumerate}

\item
The first one was based on thespectrograph SP-124 in the Nasmyth-1 focus
with a Photometrics 1K$\times$1K CCD
detector (PM1024) with $24\times24\mu$m pixel size.
We used the gratings either with 300 grooves/mm or with 600
grooves/mm. A long slit of 1\farcs5--3\farcs0$\times$40\arcsec\ was used.
The scale along the slit was 0.4\arcsec/pixel or 0.5$\arcsec$/pixel.
Various spectral set-ups were used with dispersions from 2.4 to
5.5~\AA/pixel and a wavelength coverage of 4500--7000 \AA. More details
on this set-up observations are given, e.g., in Pustilnik et
al.~(\cite{Pustilnik99}).

\item
The second set-up was based on the spectrograph SP-124 in Nasmyth-1 focus
with a Russian ISD017A 1040$\times$1160 CCD detector with $16\times16\mu$m
pixel size and quantum efficiency $\approx$50\% near $\sim$6000\AA.
This set-up was used only during 1 run in summer 1998, and few
spectra were obtained.
The same gratings and long slit were used as for the previous set-up.

\item
The most recent observations were conducted with the spectrograph LSS
(Afanasiev et al. \cite{Afanasiev95}) in the prime focus and PM1024 CCD
(as at first set-up) as a detector.
Most of the long-slit spectra (1\farcs2--2\farcs0$\times$180\arcsec) were
obtained with the grating of 325 grooves/mm, giving a dispersion
of 4.6~\AA/pixel (the spectral resolution 12-15~\AA\ (FWHM)) and
the spectral range of 3600--8000 \AA.
The scale along the slit was 0.39\arcsec/pixel.
\end{enumerate}

All observations were followed by recording the reference spectrum of a
He-Ne-Ar
lamp. Bias, dark noise and flat field were obtained every night. Observations
of the spectrophotometrical standards from 
Bohlin (\cite{Bohlin96}) were used
to derive the sensitivity curve of the overall system.
All observations and data acquisition were conducted
using the {\tt NICE} software package by Kniazev \& Shergin (\cite{Kniazev95})
in the MIDAS\footnote{MIDAS is an acronym for the European Southern
Observatory package --- Munich Image Data Analysis System.}
environment.

Since these observations were performed mainly as a back-up program, in most
of the cases the conditions were not photometric. Our main goal was to
get radial velocity of the studied candidates. Therefore we discuss
only this parameter.

Reduction was done as follows.
Cosmic ray hits removal was done in MIDAS.
The standard procedures of debiasing,
flatfielding, wavelength and flux calibration were done in
IRAF\footnote{IRAF is distributed by
National Optical Astronomical Observatories, which is operated by the
Association of Universities for Research in Astronomy, Inc., under
cooperative agreement with the National Science Foundation}.
Standard routines IDENTIFY, REIDENTIFY, FITCOORD, TRANSFORM were used
to do the
wavelength calibration and the correction for distortion and tilt for
each frame. Then the one-dimensional spectra were extracted from each
frame using the APALL routine without weighting.
To derive the
instrumental response function, we fitted the observed spectral
energy distribution of the standard stars with a high-order
polynomial.

Final measurements of the line intensities and
positions and radial velocities were done in MIDAS.
To improve the accuracy of the redshift determination,
and further, to reduce possible small systematic shifts in the
zero point of the wavelength calibration, we additionally checked the
wavelengths of the night sky emission lines on the 2D spectra at the
position of the object spectrum.

\subsection{Results of spectroscopy}

The objects observed are listed in Table A1 containing
the following information: \\
 {\it column 1:} The object's IAU-type name. \\
 {\it column 2:} Right ascension (R.A.) for equinox B1950. \\
 {\it column 3:} Declination for equinox B1950.  \\
 {\it column 4:} Apparent B-magnitude from APM database
		 (Irwin \cite{Irwin98}) which was recalculated to standard
		 CCD B-magnitude using the calibration suggested by Kniazev et al.
		 (\cite{Kniaz_HSS}). Its r.m.s. uncertainty is 0\fm45 over
		 the magnitude range B=14\fm0 to 18\fm5.\\
 {\it column 5:} Heliocentric velocity and its r.m.s. uncertainty in km~s$^{-1}$. \\
 {\it column 6:} Absolute B-magnitude calculated from the apparent B magnitude
and the heliocentric velocities. No correction for galactic extinction
is made because all observed objects are located at high galactic latitudes
and because the corrections are significantly smaller than the uncertainties
of the magnitudes. \\
 {\it column 7:} Preliminary spectral classification type according to
the presented spectral data. BCG/H{\sc ii} means that the galaxy
possesses a characteristic H{\sc ii}-region spectrum and a low enough
luminosity {\bf (M$_B \geq -$20)}. SBN are spiral galaxies of lower
excitation with the central SF
burst and the corresponding position in the line ratio diagrams, as
discussed, e.g., in Ugryumov et al. (\cite{Ugryumov99}).
Seyfert galaxies are separated mainly on the diagnostic diagrams as AGN.
The criterion of broad lines was also used for the Sy classification.
The ELG type means that an object has emission lines but is difficult
to be classified using the existing data.
ABS means a galaxy with the detected and identified absorption lines.\\
 {\it column 8:}  One or more alternative names
according to the information from NED. \\
 {\it column 9:}  The list of spectral lines, well detected in the object
spectrum and used for classification and/or redshift measurement..

All observed spectra are shown in Figures~\ref{figA1}-\ref{figA3}.

  \begin{table*}[htbp]
  \setcounter{qub}{0}
  \centering
  \begin{sideways}
  \begin{tabular}{rlllrrrlll} \\
  \MC{10}{l}{Table A1: Parameters of Observed Galaxies}\\ \hline \hline
  \#&
  \MC{1}{c}{ Galaxy }            &
  \MC{1}{c}{ $\alpha_{1950.0}$ } &
  \MC{1}{c}{ $\delta_{1950.0}$ } &
  \MC{1}{c}{ $m_{B}$ }           &
  \MC{1}{c}{ $V_{hel}$ }         &
  \MC{1}{c}{ $M_{B}^a$ }         &
  \MC{1}{c}{ Type }              &
  \MC{1}{c}{ Other }             &
  \MC{1}{c}{Identified}          \\

			       &
  \MC{1}{c}{ Name }              &
  \MC{1}{c}{ $h$~$m$~$s$ }       &
  \MC{1}{c}{ $\circ$~$\prime$~$\prime\prime$ } &
  \MC{1}{c}{ $mag$ }             &
  \MC{1}{c}{ $km/s$ }            &
  \MC{1}{c}{ $mag$ }             &
  \MC{1}{c}{       }             &
  \MC{1}{c}{ Names }             &
  \MC{1}{c}{ Lines }             \\

  &
  \MC{1}{c}{ (1) } &
  \MC{1}{c}{ (2) } &
  \MC{1}{c}{ (3) } &
  \MC{1}{c}{ (4) } &
  \MC{1}{c}{ (5) } &
  \MC{1}{c}{ (6) } &
  \MC{1}{c}{ (7) } &
  \MC{1}{c}{ (8) } &
  \MC{1}{c}{ (9) }  \\
  \hline
  \multicolumn{10}{c}{\bf SBS BCGs and discovered real neighbours}\\
  \hline
  \\[-0.15cm]
  \qq& 0916+5417  & 09 16 33.9 & +54 17 21 &+16.6 &  3640$\pm$46  &--16.8  & BCG?&               & [OII] $\lambda$3727, [OIII] $\lambda$4959+5007, H$\alpha$ \\
  \qq& 1120+5838  & 11 20 56.2 & +58 38 39 &*18.0 & 11109$\pm$39  &--17.9  & BCG?&               & H$\beta$, [OIII] $\lambda$4959+5007 \\
  \qq& 1242+5458  & 12 42 24.4 & +54 58 48 &+18.6 &  4598$\pm$60  &--15.3  & BCG?&               & [OII] $\lambda$3727, H$\beta$, [OIII] $\lambda$4959+5007, H$\alpha$ \\
  \qq& 1317+5219  & 13 17 44.1 & +52 19 24 &*16.6 &  4737$\pm$30  &--17.4  & BCG &               & H$\beta$, [OIII] $\lambda$4959+5007, H$\alpha$, [SII] \\
  \qq& 1358+5529  & 13 58 49.2 & +55 29 48 &*16.5 &  3907$\pm$27  &--17.1  & BCG &  SBS 1358+554E& H$\beta$, [OIII] $\lambda$4959+5007, H$\alpha$\\
  \qq& 1413+4934  & 14 13 19.2 & +49 34 39 &+16.8 &  3715$\pm$101 &--16.7  & BCG &  SBS 1413+495 & H$\beta$, [OIII] $\lambda$4959+5007, H$\alpha$ \\
  \qq& 1502+5307  & 15 02 31.8 & +53 07 27 &:14.7 &  5887$\pm$78  &--19.8  & ABS &  UGC 9688     & Ca{\sc ii}, G, H$\beta$, MgI, NaD, H$\alpha$ \\
  \qq& 1503+5121  & 15 03 40.0 & +51 21 06 &:15.6 &  3855$\pm$79  &--18.0  & SBN &  UGC 9702     & H$\alpha$ \\
  \qq& 1620+5744  & 16 20 15.4 & +57 44 45 &+16.8 &  4848$\pm$53  &--17.3  & BCG &  SBS 1620+577 & H$\beta$, [OIII] $\lambda$4959+5007 ,H$\alpha$, [SII] \\
  \hline
  \multicolumn{10}{c}{\bf Background galaxies}\\
  \hline
  \\[-0.15cm]
  \setcounter{qub}{0}
  \qq& 0749+5651  & 07 49 23.2 & +56 51 28 &*16.9 & 14928$\pm$158 &--19.6  & ELG   &               & [OII] $\lambda$3727, H$\alpha$, [NII] $\lambda$6584, [SII] \\
  \qq& 0749+5653  & 07 49 09.9 & +56 53 08 &*15.9 & 14136$\pm$110 &--20.5  & ABS+ELG &  MCG +10-12-007& Ca{\sc ii}, G, H$\beta$, MgI, Ca,Fe 5269\\
  \qq& 0751+5642  & 07 51 40.3 & +56 42 40 &*16.8 &  8436$\pm$306 &--18.5  & ABS   &  MCG +09-13-092& Ca{\sc ii}, G, H$\beta$, MgI, Ca,Fe 5269\\
  \qq& 0751+5650  & 07 51 48.0 & +56 50 36 &*16.7 & 13840$\pm$135 &--19.6  & ABS   &  NGC 2458     & Ca{\sc ii}, G, H$\beta$, MgI \\
  \qq& 0752+5649  & 07 52 28.9 & +56 49 14 &*16.4 &  8681$\pm$90  &--18.9  & ABS+ELG &  NGC 2462     & Ca{\sc ii}, G, H$\beta$, MgI, Ca,Fe 5269, [OIII] $\lambda$5007 \\
  \qq& 1036+4910  & 10 36 37.5 & +49 10 51 &*15.5 &  8687$\pm$11  &--19.8  & ELG   &  MCG08-20-009 & H$\alpha$, [NII] $\lambda$6584, [SII] \\
  \qq& 1037+4932  & 10 37 56.1 & +49 32 41 &*17.5 & 13051$\pm$149 &--18.7  & ELG   &  HS 1037+4932 & [OIII] $\lambda$5007, H$\alpha$, [NII] $\lambda$6584, [SII] \\
  \qq& 1037+4948  & 10 37 02.0 & +49 48 49 &*15.7 & 13457$\pm$191 &--20.6  & Sy?   &  M08-20-012   & [OII] $\lambda$3727, [OIII] $\lambda$4959+5007 ,H$\alpha$, [NII] $\lambda$6584, [SII] \\
  \qq& 1039+4938  & 10 39 35.2 & +49 38 44 &*17.0 & 13386$\pm$84  &--19.3  & ABS   &  NPM1+49.0176 & G, H$\beta$, MgI, NaD, H$\alpha$ \\
  \qq& 1130+4933  & 11 30 33.7 & +49 33 45 &*15.6 &  3191$\pm$68  &--17.5  & BCG   &  UGC 6538     & H$\beta$, [OIII] $\lambda$4959+5007, H$\alpha$, [SII] \\
  \qq& 1222+6117  & 12 22 36.1 & +61 17 25 &*16.9 & 18948$\pm$78  &--20.1  & ABS   &               & H$\beta$, MgI, NaD, H$\alpha$\\
  \qq& 1342+5923  & 13 42 49.1 & +59 23 58 &*18.3 & 21900$\pm$36  &--19.0  & ELG   &               & [OII] $\lambda$3727, H$\beta$, [OIII] $\lambda$4959+5007 ,H$\alpha$ \\
  \qq& 1401+4902  & 14 01 44.9 & +49 02 14 &+17.1 & 25360$\pm$240 &--20.5  & ABS   &  NPM1G+49.0268& H$\beta$, MgI, Ca,Fe 5269, NaD 5890-5896\\
  \qq& 1401+4907  & 14 01 20.2 & +49 07 17 &*17.1 & 22585$\pm$108 &--20.3  & ABS   &               & Ca,Fe 5269, NaD 5890-5896\\
  \qq& 1430+5242  & 14 30 48.6 & +52 42 22 &*17.6 & 13405$\pm$27  &--18.7  & BCG   &               & H$\beta$, [OIII] $\lambda$4959+5007, H$\alpha$\\
  \qq& 1448+5233  & 14 48 06.9 & +52 33 36 &*16.7 &  8990$\pm$18  &--18.7  & SBN   &               & H$\beta$, [OIII] $\lambda$5007, H$\alpha$, [NII] $\lambda$6584\\
  \qq& 1448+5240  & 14 48 24.6 & +52 40 28 &*17.6 & 28437$\pm$30  &--20.3  & Sy?   &               & [OIII] $\lambda$4959+5007 \\
  \qq& 1453+5239  & 14 53 03.9 & +52 39 31 &*18.6 & 40705$\pm$51  &--20.1  & ELG   &               & H$\beta$, [OIII] $\lambda$4959+5007 \\
  \hline \\[-0.15cm]
  \multicolumn{10}{l}{$^a$ Hubble constant $H_{0}$ = 75~km s$^{-1}$Mpc$^{-1}$ is accepted}\\
  \multicolumn{10}{l}{{\bf Notes:} {\bf 0916+5417} --- companion of BCG SBS 0916+543, which is not in the studied BCG sample;}\\
  \multicolumn{10}{l}{{\bf 1120+5838} --- companion of BCG SBS 1120+586, which is not in our sample;}\\
  \multicolumn{10}{l}{{\bf 1242+5458} --- companion of SBS 1242+549; {\bf 1317+5219} --- companion of BCG SBS~1317+523A;}\\
  \multicolumn{10}{l}{{\bf 1358+5529} --- BCG SBS~1358+554E itself; {\bf 1413+4934} --- BCG SBS~1413+495 itself;}\\
  \multicolumn{10}{l}{{\bf 1502+5307} --- companion of BCG SBS 1503+531; {\bf 1503+5121} --- companion of BCG SBS 1504+514N; }\\
  \multicolumn{10}{l}{{\bf 1620+5744} --- BCG SBS~1620+577 itself }\\
  \end{tabular}
  \end{sideways}
  \end{table*}

 \begin{figure*}[h]
 \vspace*{-0.25cm}
 \psfig{figure=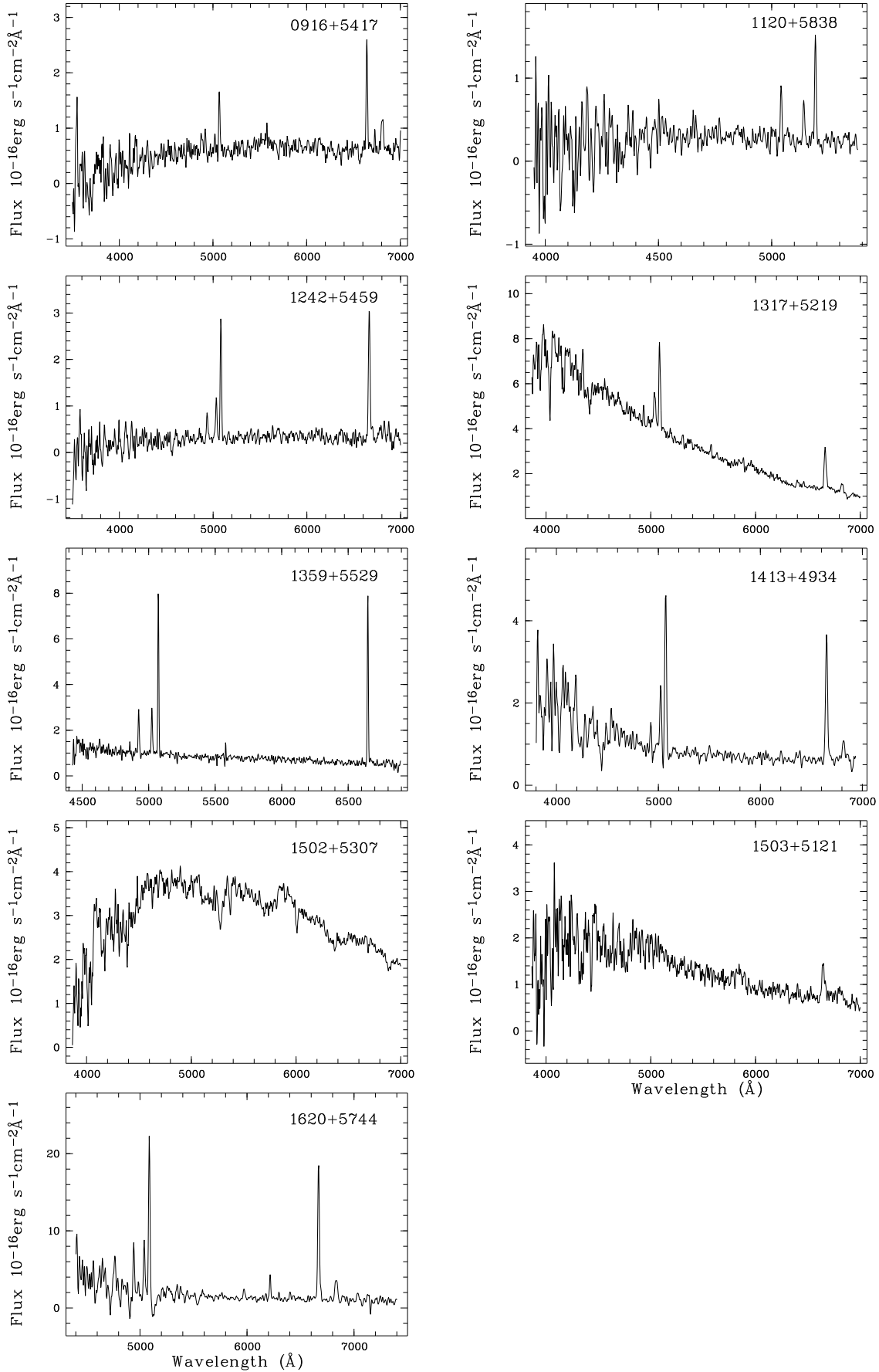,width=18.5cm,angle=0}
 \vspace*{-1.5cm}
 \centering
 \caption{Spectra of observed SBS BCGs and discovered real neighbours}
 \label{figA1}
 \end{figure*}

 \begin{figure*}[h]
 \vspace*{-0.25cm}
 \psfig{figure=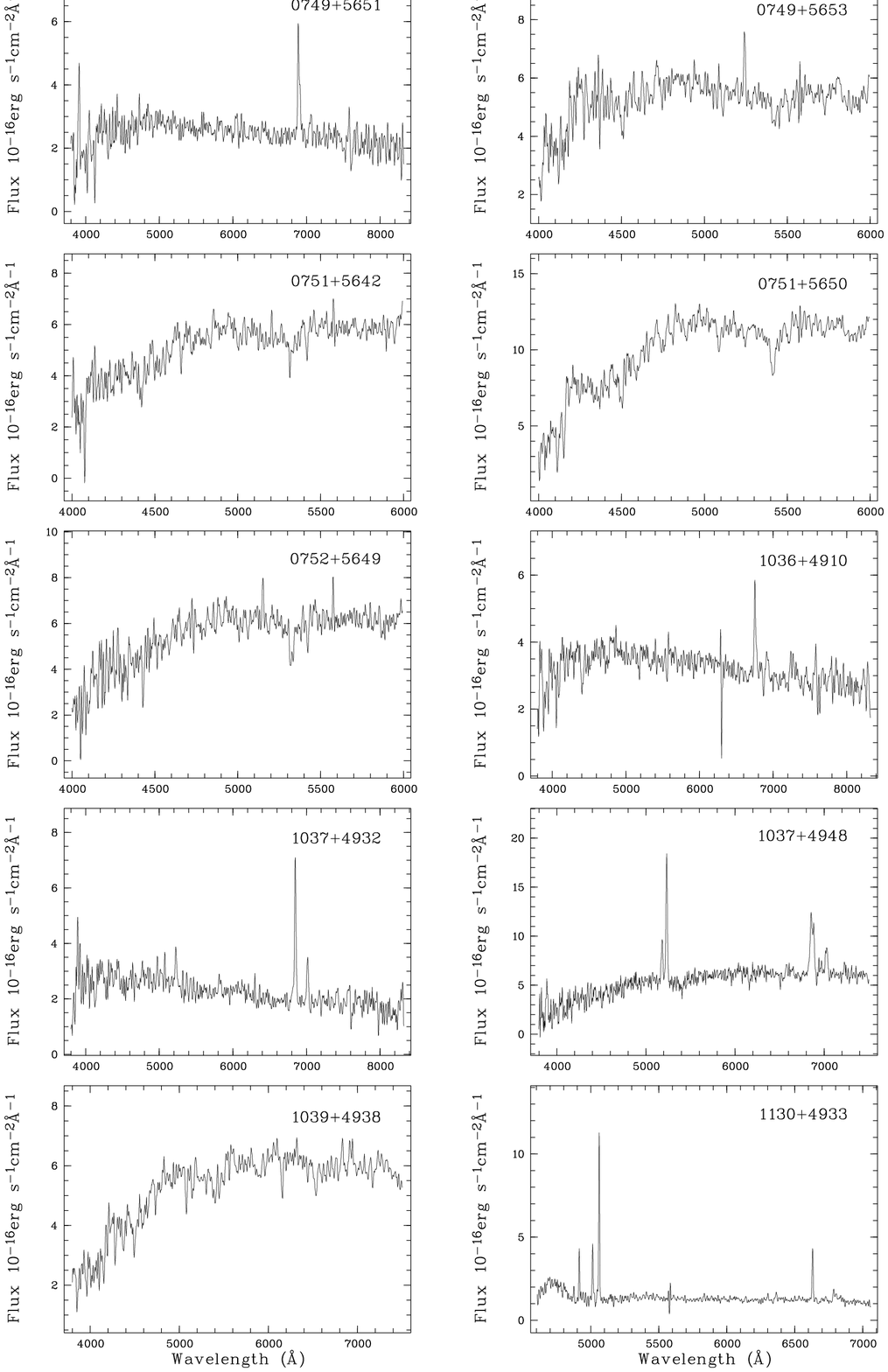,width=18.5cm,angle=0}
 \vspace*{-1.5cm}
 \centering
 \caption{Spectra of observed background galaxies}
 \label{figA2}
 \end{figure*}

 \begin{figure*}[h]
 \vspace*{-0.25cm}
 \psfig{figure=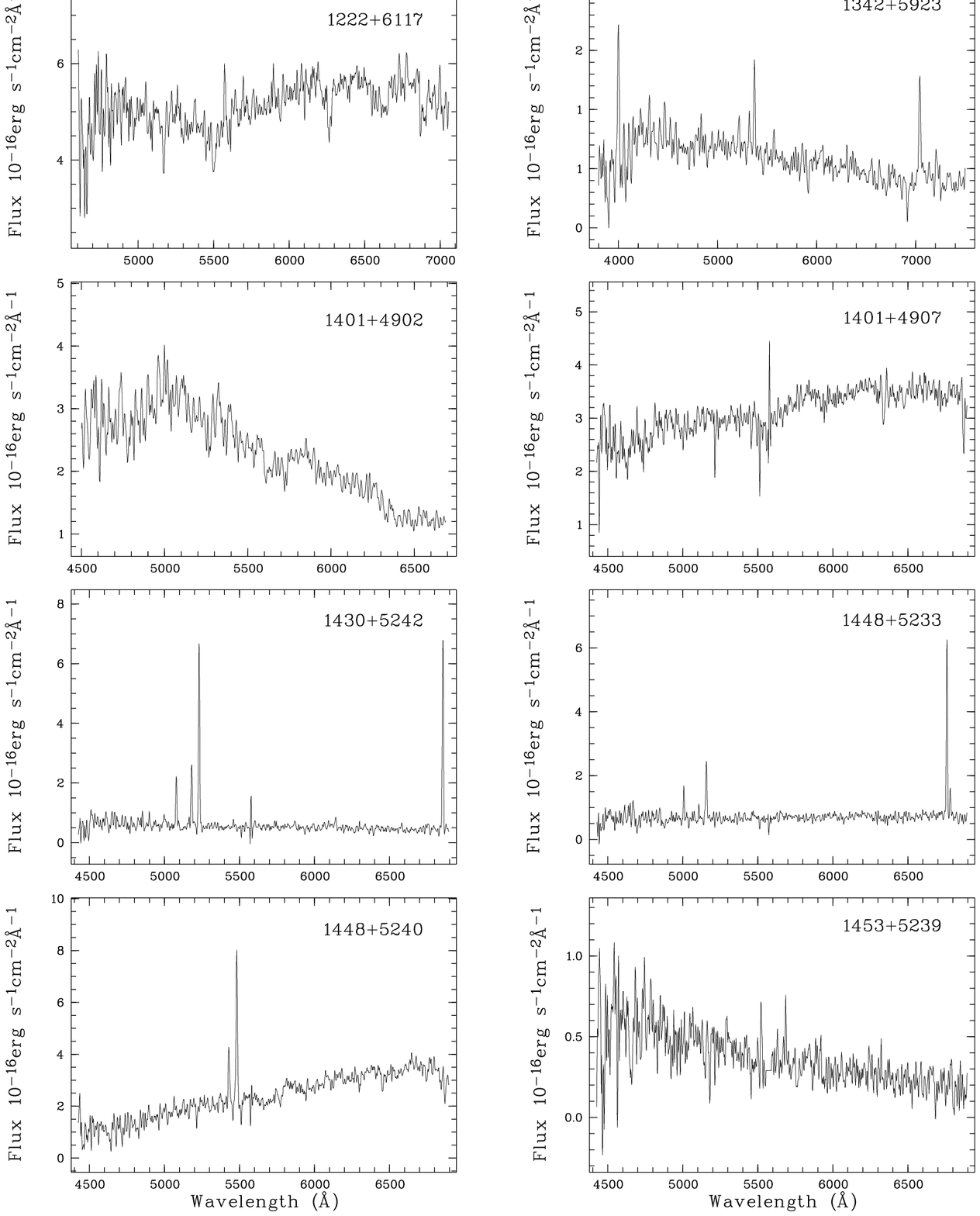,width=18.5cm,angle=0}
 \vspace*{-1.5cm}
 \centering
 \caption{Spectra of observed background galaxies}
 \label{figA3}
 \end{figure*}

\end{document}